\documentstyle[aps,prb]{revtex}
\begin{document}
\draft

\title{Theory of Fermion liquids}
\author{H.-J. Kwon, A. Houghton, and J. B. Marston}
\address{Department of Physics, Brown University, Providence, RI 02912-1843}
\date{January 24, 1995}
\maketitle

\begin{abstract}
We develop a general theory of fermion liquids in spatial dimensions greater
than one.  The principal method, bosonization, is applied to the
cases of short and long range longitudinal interactions, and to transverse
gauge interactions.  All the correlation functions of the system may be
obtained with the use of a generating functional.
Short-range and Coulomb interactions do not destroy the Landau Fermi fixed
point.  Novel fixed points are found, however, in the cases of a super-long
range longitudinal interaction in two dimensions and transverse
gauge interactions in two and three spatial dimensions.
We consider in some detail the 2+1-dimensional problem of a Chern-Simons
gauge action combined with a longitudinal two-body interaction
$V({\bf q}) \propto |{\bf q}|^{y-1}$ which controls the density, and hence
gauge, fluctuations.  For $y < 0$ we find that the gauge interaction is
irrelevant and the Landau fixed point is stable, while for $y > 0$
the interaction is relevant and the fixed point cannot be
accessed by bosonization.  Of special importance is the case
$y = 0$ (Coulomb interaction) which describes the
Halperin-Lee-Read theory of the half-filled Landau level.  We obtain
the full quasiparticle propagator which is of a marginal Fermi liquid
form.  Using Ward Identities, we show that neither the inclusion of
nonlinear terms in the fermion dispersion, nor vertex corrections, alters
our results: the fixed point is accessible by bosonization.
As the two-point fermion Green's function is not gauge
invariant, we also investigate the gauge-invariant density response
function.  Near momentum $Q = 2 k_F$, in addition to the
Kohn anomaly we find singular
behavior.  In Appendices we present a numerical calculation of the
spectral function for a Fermi liquid with Landau
parameter $f_0 \neq 0$.  We also show how Kohn's theorem is satisfied
within the bosonization framework.

\end{abstract}


cond-mat/9501067

\medskip
\section{Introduction}
\label{sec:intro}
The peculiar nature of electrons in a half-filled Landau level, and anomalies
in the normal state transport properties of the high-temperature
cuprate superconductors, have stimulated great interest in systems of
strongly interacting fermions moving in spatial dimensions greater than one.
In addition to methods such as mean-field theory\cite{Halperin},
RPA\cite{Halperin,Furusaki}
and 1/N perturbation theory\cite{Furusaki,Altshuler},
a number of non-perturbative
theoretical tools such as the renormalization group,\cite{Shankar,Gan,Nayak}
the Eikonal approximation,\cite{Khvesh} and Ward identities\cite{Castellani}
have been used to study fermion liquids.
Bosonization\cite{Haldane,Tony,HKM,HKMS,KHM,Castro,Castro2,Kopietz,DVK}
is another powerful tool for understanding the effects of both regular and
singular interactions
on fermion liquids as the realm in which it is applicable, for low-energy
excitations near the Fermi surface, is precisely where these
interactions are most
important.  Bosonization does not rely upon a Fermi liquid
form for the quasiparticle propagator; consequently non-trivial zero
temperature quantum critical fixed
points are accessible.  In an appropriate limit, bosonization reduces
complicated many-body Hamiltonians to quadratic forms which possess infinite
$U(1)$ symmetry and which can be diagonalized exactly.

A general theory of fermion
liquids in two and three spatial dimensions requires the consideration of
a variety of Fermi surface geometries and interactions.
In this paper we study only
the simplest Fermi surface geometries: a circle in spatial dimension
$D = 2$ or a sphere in $D = 3$.  We assume that the symmetries of these
surfaces do not break
spontaneously when the interactions between fermions are turned on.
The bosonization formalism applies, without significant modification,
to more complicated Fermi surfaces provided there are no Van Hove singularities
and nesting does not occur. However, the added technical complications
of describing
these more complicated surfaces would make detailed discussion of their
properties cumbersome and, given this caveat,
would not lead to any qualitative changes in the
physics.  We do consider a variety
of two-body interactions between the fermions.  In accord with expectations,
we find, barring superconducting instabilities, that
short-range and Coulomb interactions do not destroy the Landau fixed point.
Novel fixed points are found, however, in the cases of super-long range
interactions and transverse gauge interactions.  The latter case is
particularly significant as
gauge fields play an important role in the physics of the half-filled
Landau level\cite{Halperin} and may be important in
the cuprate superconductors\cite{Gauge,Lee}.
The effects of transverse gauge fields,
which are not screened, are particularly interesting as perturbative
expansions in the coupling constant break down\cite{Blok}.

Earlier work on multidimensional bosonization showed that it is an accurate
calculational framework for understanding Landau Fermi
liquids\cite{Haldane,Tony,HKM,HKMS,KHM,Castro,Castro2,Kopietz,DVK}.
Here we address the question of whether the
approximations inherent in bosonization
are adequate for the method to apply to
a wider range of fermion liquids with singular or transverse
gauge interactions.  Three questions in particular must be
answered in order to check that bosonization does not break down.
First, the assumed current algebra must be consistent: it is derived for the
free system, and it must continue to hold when interactions are turned on.
Second, we examine whether or not the neglect of quadratic terms
in the fermion quasiparticle spectrum due to Fermi surface curvature is
an innocuous approximation.  Finally, we need to check whether certain vertex
corrections are significant by returning to the fermion basis to examine
their contribution.  In this paper, we examine systems which meet all
three of the above conditions.  The fixed points of these systems therefore
can be accessed and studied via bosonization.

The outline of this paper is as follows.
In Sec. \ref{sec:bose}, we develop a bosonization formalism applicable in
any spatial dimension which
describes both longitudinal and transverse gauge interactions.  A
generating functional permits the easy computation of 2N-point boson and
fermion correlation functions.  Interactions of interest are simply
special cases of this general formalism.
In Sec. \ref{sec:longit}, we demonstrate how to incorporate any two-body
density-density interaction by introducing
longitudinal gauge fields which mediate the interaction.
As examples, we reproduce the Luttinger liquid in one spatial dimension and
solve the problem of a single, but general, Landau parameter
$f_n$ in two and three spatial dimensions.
In Appendix A we use the numerical method of fast-Fourier-transforms (FFT)
to extract the spectral function of a two-dimensional Fermi liquid with
Landau parameter $f_0$.  And in Appendix B we present details of our solution
for more complicated Landau parameters.  Then we turn to the problem
of long-range longitudinal interactions.  We show that
screening of the Coulomb interaction occurs naturally within the bosonization
framework. By considering fermions with spin, we show that
spin-charge separation does not occur in $D > 1$.
In the case of super-long range interactions in $D=2$,
however, the Landau fixed point is destroyed by the emission of plasmons by
the quasiparticles.  We show that bosonization accurately describes this
breakdown.

The fact that bosonization can access non-Fermi liquid
fixed points, both in one and two spatial dimensions, motivates
Sec. \ref{sec:transverse}, in which we study a more
physically relevant example of the breakdown of Fermi liquid behavior: the
Chern-Simons theory of half-filled Landau level put forward by Halperin, Lee,
and Read\cite{Halperin} (HLR) and Kalmeyer and Zhang\cite{Kalmeyer}.
To investigate the range of validity of bosonization, we examine the general
case of an interaction of the form $V({\bf q}) = |{\bf q}|^{y-1}$
with $-1 < y \leq 1$ when the inverse gauge propagator is given by
$i\gamma |\omega |/|{\bf q}|-\chi |{\bf q}|^{y+1}$.
Aided by the Ward Identity approach, we find that
for $y < 0$ the gauge interaction is irrelevant
and the Landau fixed point is stable, while for $y > 0$ the interaction
is relevant and the fixed point cannot be directly studied by bosonization.
We discuss in detail the special case $y = 0$ corresponding to the Coulomb
interaction.  In this case a marginal Fermi liquid (MFL)
fixed point controls the low energy physics.
(An brief account of this work has appeared in print\cite{KHM}.) However,
as the two-point Green's function is not gauge invariant, it is not directly
related to experimentally observable quantities.  The density
response function, on the other hand, is gauge invariant and in
Sec. \ref{sec:dens} we study its behavior for $Q \approx 0$ and $Q \approx
2k_F$.  In the former case our results are identical to those found in
RPA\cite{Furusaki} while in the latter case we find non-analytic
structure stronger than that found by Altshuler, Ioffe and
Millis\cite{Altshuler}.  In Sec. \ref{sec:EM}, we apply the formalism developed
in Secs. \ref{sec:bose} and \ref{sec:transverse}
to the case of the physical Maxwell electromagnetic interaction in three
dimensions to obtain the
quasiparticle Green's function in the Coulomb gauge.  Although the Landau fixed
point is unstable at extremely low energies,
in accord with expectations it controls
the behavior at physically relevant temperatures.
A brief summary of our work and conclusions are found in Sec. \ref{sec:conc}.
In Appendix C, we include the Landau parameter $f_1$ in the HLR theory to show
that Kohn's theorem\cite{Kohn} is recovered automatically: the bare
cyclotron frequency, not a renormalized value, appears in the collective mode
spectrum.

\section{Bosonization}
\label{sec:bose}

In this section we formulate the problem of an interacting fermion
liquid, making use of a bosonization method applicable in any spatial
dimension to study both longitudinal and transverse gauge
interactions. A generating functional permits the ready computation of
2N-point boson and fermion correlation functions.

We start from the bare Hamiltonian for fermions interacting with non-compact
longitudinal and transverse gauge fields $A^{\mu }$ in $D$-spatial
dimensions.  For simplicity, we consider spinless fermions and spherical
Fermi surfaces only. In the Coulomb gauge, ${\bf \nabla \cdot A}=0$,
and when $\hbar $ is set equal to one,
\begin{equation}
H = \int d^Dx ~c^{\dag}({\bf x })\big{[}{(-i{\bf \nabla -A})^2\over 2m}
- A_0 \big{]}c({\bf x}) +{{1}\over{2}}~
\int d^Dx~ d^Dy~ V({\bf x-y})~c^{\dag}({\bf x})~ c^{\dag}({\bf y})~
c({\bf y})~ c({\bf x})
\label{bare}
\end{equation}
where $\epsilon_{\bf k} \equiv {\bf k}^2/2m$,
and $V({\bf x-y})$ represents Coulomb or other two-body interactions.
Next we make use of the renormalization group to integrate out the
high-energy Fermi degrees of freedom\cite{Shankar}.
The resulting low-energy effective theory is
expressed in terms of quasiparticles $\psi_{\bf k} $ which obey
canonical anticommutation relations and which are
related to the bare fermion operators by
\begin{equation}
\psi_{\bf k} = Z^{-1/2}_{\bf k}~ c_{\bf k}
\label{ZK}
\end{equation}
for momenta $\bf k$ which are restricted to a narrow shell of thickness
$\lambda$
around the Fermi surface: $k_F - \lambda/2 < |{\bf k}| < k_F + \lambda/2$.
The wavefunction renormalization factor
$Z_{\bf k}$ rescales the discontinuity in the quasiparticle
occupancy at the Fermi surface back to one.
To be explicit, consider the partition function
\begin{equation}
Z = \int {\cal D}A^\mu ~{\cal D}\psi ~{\cal D}\not\!\psi ~e^{i~S[A,~ \psi,~
\not\!\psi]}
\end{equation}
where we have separated the fermion fields into low and high energy components,
$\psi({\bf x})$ and $\not\!\psi ({\bf x})$ respectively.  Integrating out the
high energy $\not\!\psi$ fields, which contain all of the degrees of freedom
except for the narrow shell around the Fermi surface,
we obtain the effective action:
\begin{eqnarray}
S[A, \psi] &=&\int d^Dx~dt ~\big{[}\psi^{*}({\bf x })(i\partial _t +A_0)
\psi({\bf x})
+{1\over 2m^*}\psi^{*}({\bf x})({\bf \nabla}-i~{\bf A})^2 \psi({\bf x})\big{]}
\nonumber \\
&&+ \sum_{\bf q}~ \int {{d\omega}\over{2 \pi}}~
\not\!\Pi_{\mu \nu}(q)~ A^\nu(q)~ A^\mu(-q)~ .
\label{FL}
\end{eqnarray}
We use the notation $q \equiv (\omega, {\bf q})$.
Renormalization of the Fermi velocity
$v_F \rightarrow v_F^*$ is expressed through the appearance of the
effective mass $m^*$.
The last term in Eq. (\ref{FL}), which
is generated by integrating out the
$\not\!\psi$ fields deep inside the Fermi sea contributes to the
total diamagnetic term,
$-(\rho_f / 2 m^*) {\bf A}(q) {\bf \cdot} {\bf A}(-q)$; $\rho_f$ is the fermion
number density.  Irrelevant operators consisting of higher powers and/or
derivatives of the gauge and Fermi degrees of freedom are also
generated.  These operators may be safely neglected in the
low-energy limit assuming that the anomalous dimension of the gauge fields
\cite{Gan,Nayak} is not too different from their engineering dimension.
To check that this is the case, the effective action of the gauge fields must
be obtained by integrating out the fermion degrees of freedom\cite{Gan,Hertz},
and examined for the case of interest.
The effects of two-body fermion interactions which in general include
both the long range Coulomb interaction and short range Fermi liquid type
interactions will be considered later in the discussion.

To bosonize the effective action Eq.(\ref{FL}) we introduce the coarse
grained charge current\cite{Tony}
\begin{equation}
J({\bf S; q}) \equiv \sum_{\bf k} \theta({\bf S; k + q})~
\theta({\bf S; k})~ \{ \psi^{\dagger}_{\bf k + q}~
\psi_{\bf k} - \delta^3_{\bf q, 0}~ n_{\bf k} \}\ .
\label{curk}
\end{equation}
Here ${\bf S}$ labels a patch [${\bf S}\equiv (\theta , \phi )$ in
three dimensions] on the Fermi surface with momentum ${\bf k_S}$, and
$\theta({\bf S; k}) = 1$ if ${\bf k}$ lies inside a squat box
centered on $\bf S$ with height $\lambda$ in the radial (energy) direction
and area $\Lambda^{D-1}$ along the Fermi surface, and equals zero
otherwise. These two scales must be small in the sense
$k_F \gg \Lambda \gg \lambda \geq |{\bf q}|$:
we satisfy these limits by setting
$\lambda \equiv k_F/N$ and $\Lambda \equiv k_F/N^\alpha$ where
$0 < \alpha < 1$ and $N \rightarrow \infty$.
The currents so defined satisfy the $U(1)$ current algebra
\begin{equation}
[J({\bf S; q}), J({\bf T; p})]=\delta ^{D-1}_{\bf S,T}
\delta ^{D}_{\bf q+p,0} \Omega ~{\bf q\cdot \hat{n}_S} ~ .
\label{u1alge}
\end{equation}
Here $\Omega \equiv \Lambda^{D-1} (L/2 \pi)^D$ is the number of states
in the squat box divided by $\lambda$ and ${\bf \hat{n}_S}$ is a unit
normal to the Fermi surface at ${\bf S} $. The charge currents also
have a bosonic representation; if we set
\begin{equation}
J({\bf S; x}) = \sqrt{4\pi } {\bf \hat{n}_S \cdot \nabla \phi({\bf S; x})}
\end{equation}
where the bosonic charge fields satisfy the canonical commutation relations,
\begin{eqnarray}
[\phi({\bf S; x}),\phi({\bf T; y})]
&=& {i\over 4}\Omega ^2 \delta^{D-1}_{\bf S,T}~
\epsilon \Big{(} {\bf \hat{n}_S \cdot[x-y]}\Big{)}~;~
|x_{\perp}-y_{\perp}|\Lambda \ll 1 \nonumber \\
&=& 0 ~ ; ~|x_{\perp}-y_{\perp}|\Lambda \gg 1
\end{eqnarray}
where $\epsilon(x) = 1$ for $x > 0$ and equals $-1$ otherwise.
Eq. (\ref{u1alge}) follows immediately from the commutation relations.
The key formula of the bosonization procedure expresses the fermion
quasiparticle field $\psi $ in terms of the boson fields as
\begin{equation}
\psi({\bf S; x}, t) = {1\over\sqrt{V}}~ \sqrt{{\Omega}\over{a}}
e^{i{\bf k}_{\bf S}{\bf \cdot x}}
\exp \{i{\sqrt{4\pi}\over \Omega} \phi({\bf S; x}, t)\}~ \hat{O}({\bf S})~.
\label{bosonization}
\end{equation}
Here $V$ is the volume of the system and $a \equiv 1/\lambda$ is an ultraviolet
cutoff.  $\hat{O}(S)$ is an ordering operator introduced\cite{Tony,Luther}
to maintain Fermi statistics in the angular direction along
the Fermi surface.  Anticommuting statistics are obeyed automatically
in the direction normal to the Fermi
surface, just as in one-dimensional bosonization.
It is now an interesting exercise to use this representation of
the fermion field operators to verify that
\begin{eqnarray}
J({\bf S; x}) &=& V \lim_{\epsilon \rightarrow 0}
\big{\{} \psi^{\dag}({\bf S; x+\epsilon \hat{n}_S})~
\psi({\bf S; x}) - \langle \psi^{\dag}({\bf S; x+\epsilon \hat{n}_S})~
\psi({\bf S; x}) \rangle \big{\}}
\nonumber \\
&\equiv & \sqrt{4\pi} {\bf \hat{n}_S\cdot \nabla } \phi({\bf S; x}) ~ .
\end{eqnarray}
The currents are invariant under $U(1)$ phase rotations through an
angle $\beta({\bf S})$ in each patch:
$\psi({\bf S; x}) \rightarrow e^{i \beta({\bf S})}~ \psi({\bf S; x})$.
This infinite $U(1)$ symmetry reflects the fact that the current operator
in a given patch does not scatter quasiparticles outside of that patch.
The effective action Eq. (\ref{FL}) now can be expressed in terms of
the charge currents $J$ and consequently reflects the underlying infinite
$U(1)$ symmetry of the fixed point:
\begin{eqnarray}
S[A,a] &=& \sum_{\bf S} \sum_{\bf q, q\cdot \hat{n}_S >0}
\int {d\omega \over 2\pi}
{}~(\omega - v_F^*~ {\bf q\cdot \hat{n}_S})~a^{*}({\bf S}; q)~a({\bf S}; q)
\nonumber \\
&+& {1\over V} \sum_{\bf q }\int {d\omega \over 2\pi }
\big{\{} \sum_{\bf S} J({\bf S}; q)[ A_0(-q) +
{\bf v_S^*\cdot A}(-q)] -{\rho_f\over 2 m^*} {\bf A}(q)\cdot {\bf A}(-q)
\big{\}} ~.
\label{action}
\end{eqnarray}
Here ${\bf v_S^* \equiv k_S} / m^*$ is the renormalized Fermi velocity vector
at patch $\bf S$ and
the charge currents are related to the canonical boson operators $a$ and
$a^\dagger$ by:
\begin{equation}
J({\bf S; q}) = \sqrt{ \Omega~ |{\bf \hat{n}_S \cdot q}|}~
[ a({\bf S; q})~ \theta({\bf \hat{n}_S \cdot q}) + a^\dagger({\bf S; -q})~
\theta(-{\bf \hat{n}_S \cdot q})]\ , \label{canon}
\end{equation}
where
\begin{equation}
[a({\bf S;q}),~a^{\dag}({\bf T;p})]=\delta^{D-1}_{\bf S,T}~\delta^D_{\bf q,p}
\end{equation}
and $\theta(x)=1$ if $x>0$ and is zero otherwise.
Exchange scattering, which involves high momentum gauge fields, has not been
included in the action, as it is not singular in the low-momentum limit.
The action Eq. (\ref{action}) must be supplemented by the bare gauge
action which we take to have the quadratic form:
\begin{equation}
S^0_G[A] = {1\over 2}\int {d^Dq\over (2\pi)^D}\int
{d\omega \over 2\pi }~ K^0_{\mu \nu}(q)~ A^\mu(q)~ A^\nu(-q)~.
\label{SG0}
\end{equation}
We fix the gauge with the supplemental condition ${\bf \nabla \cdot  A}=0$
(ie. the Coulomb gauge).  Particular examples considered in this
paper are the physical electromagnetic gauge action in $D = 3$, and
in $D=2$ the Chern-Simons action, which breaks time-reversal and parity,
but not charge-conjugation, symmetries.

We proceed with the evaluation of the boson correlation function
which can be carried out exactly.
Rather than integrate out the gauge fields to obtain an effective
current-current interaction,  we introduce a generating functional
for the boson correlation function and first
integrate out the boson fields $a$ and
$a^*$.  In this way we avoid the intermediate step of summing an infinite
perturbation series in  the effective interaction to obtain the
boson propagator\cite{HKM}.  We construct the generating functional by coupling
fields $\xi $ and $\xi^*$ to the boson fields $a^{*}$ and $ a$.
\begin{eqnarray}
Z[\xi, \xi^*] &=& \int {\cal D}^\prime A^\mu ~{\cal D}a~{\cal D}a^{*}~
\exp \{i(S[A,a]+S^0_G[A])\}~ \nonumber \\
&& \times \exp \big{\{}
i\sum_{\bf S}\sum_{\bf q, q\cdot \hat{n}_S>0}
\int {d\omega \over 2\pi }~[\xi ({\bf S}; q) a^{*}({\bf S}; q)
+ \xi^*({\bf S}; q) a({\bf S}; q)] \big{\}}~ .
\end{eqnarray}
Here the prime over the gauge measure indicates that the integration respects
the Coulomb gauge fixing condition.
Then by completing the square we integrate out the boson fields in each
patch $\bf S$.  It is important to note that the entire Fermi surface
participates: patches at every point on the Fermi surface contribute
to the effective action.
\begin{eqnarray}
Z[\xi, \xi^*]
&=& {\cal N}\int {\cal D}^\prime A^\mu ~\exp\{i~S_G[A]\}
\exp \bigg{\{} -i\sum_{\bf S}\int{d^Dq\over (2\pi)^D}~
\int {d\omega \over 2\pi }~
{{\theta({\bf q \cdot \hat{n}_S})}\over{\omega -v_F^* {\bf q\cdot \hat{n}_S}}}~
\nonumber \\
&\times& \Big{[} \sqrt{\Omega {\bf q\cdot \hat{n}_S}}~
[\xi ({\bf S}; q)~ (A_0(-q)+ {\bf v_S^* \cdot A}(-q))
+ \xi^*({\bf S}; q)~(A_0(q) +  {\bf v_S^* \cdot A}(q))] \nonumber \\
&+& V \xi({\bf S}; q)~ \xi^*({\bf S}; q) \Big{]} \bigg{\}}~ ,
\label{ftnal}
\end{eqnarray}
where $S_G[A]$ is an effective gauge action, which since the bosons were
coupled linearly to the gauge fields is a quadratic form,
\begin{equation}
S_G[A] = {1\over 2}\int {d^Dq\over (2\pi)^D}\int
{d\omega \over 2\pi }~ K_{\mu \nu}(q)~ A^\mu(q)~ A^\nu(-q)~.
\label{SG}
\end{equation}
The inverse of the gauge propagator satisfies the equation
\begin{equation}
K_{\mu \nu }(q)~ A^\mu(q)~ A^\nu(-q) =
K^0_{\mu \nu}(q)~ A^\mu(q)~ A^\nu(-q)
+ \chi ^0(q)~ A_0(q)~ A_0(-q) + \chi^T(q)~ {\bf A}(q) \cdot {\bf A}(-q)~ ,
\label{Kmn}
\end{equation}
where $\chi^0$ and $\chi^T$ are the longitudinal and
transverse susceptibilities which for the cases of perfectly circular or
spherical Fermi surfaces, are given by
\begin{eqnarray}
\chi^0(q) &=& N^*(0) \big{[} 1 -{\theta(x^2-1)~|x|
\over \sqrt{x^2-1}}+i{\theta(1-x^2)~|x|\over \sqrt{1-x^2}}\big{]}
{}~ ; ~ D=2 \nonumber \\
&=& N^*(0) \big{[} 1-{x\over 2}\ln \big{(}{x+1\over x-1}\big{)}\big{]}
{}~ ; ~ D=3 ~ ,
\label{chi0}
\end{eqnarray}
and
\begin{eqnarray}
\chi ^T(q)&=&v_F^{* 2}~ N^*(0) \big{[} -x^2 +\theta(x^2-1)~|x|\sqrt{x^2-1}
+i\theta(1-x^2)~|x|\sqrt{1-x^2} \big{]} ~ ; ~ D=2
\nonumber \\
&=&{1\over 2}v_F^{*2}~N^*(0) \big{[}-x^2-{x\over 2}(1-x^2)\ln \big{(} {x+1\over
x-1}\big{)} \big{]} ~ ; ~ D=3 ~ ,
\label{chiT}
\end{eqnarray}
where $x \equiv \omega / (v_F^*~|{\bf q }|)$, and $N^*(0)$,
which is the quasiparticle density of states at the Fermi surface, equals
$m^*/2\pi$ for spinless fermions
in $D=2$, and $m^*k_F / 2\pi^2$ in $D=3$.
The integral over the gauge fields $A^{\mu}$ can be carried out now
giving the generating functional as an explicit function of $\xi $ and
$\xi^*$. The boson correlation functions are obtained by differentiating
the logarithm of the generating functional with respect to
$\xi $ and $\xi^*$. In particular the boson propagator defined by
\begin{equation}
\langle a({\bf S}; q)~ a^{\dagger}({\bf S}; q)\rangle
= -{{\delta^2 \ln Z[\xi, \xi^*]}\over{\delta \xi({\bf S}; q)
{}~\delta \xi^*({\bf S}; q)}}|_{\xi = \xi^* = 0}
\label{derivat}
\end{equation}
is given in terms of the gauge propagator
$D_{\mu \nu}(q) = \big{[} K(q)^{-1} \big{]}_{\mu \nu}$ as
\begin{eqnarray}
\langle a({\bf S}; q)~ a^\dagger({\bf S}; q)\rangle
&=& {i\over \omega - v_F^* {\bf q\cdot \hat{n}_S} +i \eta~ {\rm sgn}(\omega)}
+ i~ {\Lambda^{D-1} \over (2\pi )^D}~ {\bf q\cdot \hat{n}_S}~
{D_{\mu \nu }({\bf q},\omega )~\epsilon ^\mu ({\bf S};q)
{}~\epsilon ^\nu ({\bf S};-q)\over
[\omega - v_F^* {\bf q\cdot \hat{n}_S} + i \eta ~{\rm sgn}(\omega)]^2} ~ ,
\label{Gaa}
\end{eqnarray}
where $\epsilon ^\mu ({\bf S};q)$ is the D+1-vector $(1, {\bf v_S^*})$.
Note that the velocity of the interacting bosons differs from that of
free bosons by at most of order
$v_F^*~ (\Lambda/k_F)^{D-1}$.  Therefore
velocity renormalization of the bosons due to interactions among
the final shell of excitations
with energy less than $v_F^* \lambda$ is insignificant in the $\Lambda
\rightarrow 0$ limit except in the special
case of one spatial dimension.  We will return to this observation
below when we address the question of whether or not
spin-charge separation occurs in two or higher spatial dimensions.

Now the bosonization formula Eq. (\ref{bosonization}) can be used
to express the space-time fermion Green's function in terms of the
Fourier transform of the boson correlation function. In general
terms the Green's function for fermions in a single patch on the
Fermi surface is given by
\begin{equation}
G_F({\bf S;x},t) = {\Omega \over V~a} e^{i{\bf k_S \cdot x}} \exp \bigg{[}
\int {d^Dq\over (2\pi )^D}\int {d\omega
\over 2\pi }~[e^{i({\bf q\cdot x}-\omega t)}-1]~{(2\pi )^D\over
\Lambda ^{D-1}~{\bf \hat{n}_S \cdot q}}~\langle a({\bf S};q)~
a^{\dag }({\bf S};q)\rangle \bigg{]} ~ ,
\label{Gff}
\end{equation}
which becomes, upon using Eq. (\ref{Gaa}) and dropping the prefactor of
$\Omega/V$ for brevity,
\begin{eqnarray}
G_F({\bf S;x},t) &=&
{e^{i {\bf k_S\cdot x}}\over {\bf x \cdot \hat{n}_S}- v_F^* t}~
\exp \bigg{[} i \int {d^Dq\over (2\pi )^D}\int {d\omega \over 2\pi }~
[e^{i({\bf q\cdot x}-\omega t)}-1]~
{D_{\mu \nu }({\bf q},\omega )~\epsilon ^\mu ({\bf S};q)
{}~\epsilon ^\nu ({\bf S};-q)\over
[\omega - v_F^* {\bf q\cdot \hat{n}_S} + i \eta ~{\rm sgn}(\omega)]^2}
\bigg{]}
\nonumber \\
&=& {e^{i {\bf k_S\cdot x}}\over {\bf x \cdot \hat{n}_S}- v_F^* t}~
\exp \bigg{[} \delta G_B({\bf S; x}, t) \bigg{]}~ .
\label{Gffc}
\label{fermi}
\end{eqnarray}
for ${\bf x_\perp \equiv x \times \hat{n}_S = 0}$.
The condition $\bf x_\perp = 0$ can be replaced by
$|{\bf x_\perp}| \Lambda < 1$ provided we control the logarithmic
infrared divergence of the free boson correlation function
by placing the system inside a large box.  The weak divergence, which is a
consequence of treating the Fermi surface as locally flat inside each
patch, presents no real difficulties in practice, and will be ignored
in the following analysis.
For $|{\bf x_\perp}| \Lambda \gg 1$ the fermion Green's function vanishes
since bosons separated by large $\bf x_\perp$ are uncorrelated.
In the second line of Eq. (\ref{Gffc}) for convenience
we have introduced the notation $\delta G_B$ for the modified
Fourier transform of the additive correction to the free boson
propagator due to interactions.
\begin{equation}
\delta G_B({\bf S; x}, t) =
i \int_{-\lambda/2}^{\lambda/2}
{d^D q\over (2\pi )^D}\int {d\omega \over 2\pi }~
[e^{i({\bf q\cdot x}-\omega t)}-1]~ \delta G_B({\bf S; q}, \omega)\ ,
\label{Gb}
\end{equation}
where we can identify $\delta G_B({\bf S; q}, \omega)$ as:
\begin{equation}
\delta G_B({\bf S; q}, \omega) =
{{D_{\mu \nu }({\bf q},\omega )~\epsilon ^\mu ({\bf S};q)
{}~\epsilon ^\nu ({\bf S};-q)}\over
{[\omega - v_F^* {\bf q\cdot \hat{n}_S} + i \eta ~{\rm sgn}(\omega)]^2}}\ .
\label{additive}
\end{equation}
In Eq. (\ref{Gb}), the momentum integral over $\bf q$ ranges over $(-\lambda/2,
\lambda /2)$ in each direction.   In contrast, in the {\it free} part of
the boson and fermion propagators, momenta in the perpendicular
directions range over the much larger interval $(-\Lambda /2,\Lambda /2)$.
The reason for the difference lies
in the fact that interactions which couple different patches on the Fermi
surface must be cutoff at momentum $\lambda \ll \Lambda$ because in general
the Fermi surface normal vector points in a different direction for each patch.
Therefore, as the patches are squat pillboxes with dimensions
$\Lambda^{D-1} \times \lambda$,
only modes with $|{\bf q}| < \lambda$ are permitted
by the geometry of the construction to couple patches together.

To complete the program of computing the fermion two point function, the
Green's function in Eq. (\ref{Gffc}) is summed over all patches on the
Fermi surface:
\begin{equation}
G_F({\bf x}, t) = {\sum_{\bf S}}^\prime~ G_F({\bf S; x}, t)\ .
\end{equation}
The prime indicates that the sum is only over patches for which
$|{\bf x \times \hat{n}}_S| \Lambda < 1$.
The expression for the fermion Green's function given here coincides with
that found by Castellani {\it et al.} \cite{Castellani} who make use
of Ward Identities derived in the fermion basis to find Eq. (\ref{Gffc}).
Why these two different approaches should give the same result and
the consequent implications are discussed in detail in Sec.
\ref{subsec:ward}.

\section{Longitudinal Interactions}
\label{sec:longit}

In this section we show how the generating functional may be used to
determine the correlation functions of a fermion liquid interacting
via two-body forces in two important special cases, for short-range
interactions of Landau Fermi liquid type, and for Coulomb or longer range
longitudinal interactions.
The procedure parallels that given in the previous section as we
introduce fictitious gauge fields, which are linearly coupled to the bosons,
to mediate  the interaction. First, to illustrate the non-perturbative
nature of the method, we determine the propagator of a fermion liquid
in one spatial dimension, which has a Luttinger liquid form.
In higher dimensions, we extend the discussion to general short-range
interactions which as expected do not destroy the Fermi liquid
state. The quasiparticle propagator for a system of fermions
interacting via a short-range interaction described by a single Landau
parameter $f_0$ had been derived in a previous work\cite{HKM} by a
different technique.
Here we give a detailed derivation of the propagator for the nontrivial
case of a fermion liquid with an $f_1$ interaction. This result will
be used later in the derivation of Kohn's theorem which is applied in
the discussion of the half-filled Landau level studied in the next section.
The derivation for a general interaction specified by more than one
Landau parameter and a proof of Kohn's theorem are given in
Appendices B and C respectively.  Finally, we show
that, due to screening, the ground state of a fermion system
interacting via the Coulomb interaction is a Landau Fermi liquid in two
and three dimensions.  For a super-long-range interaction in $D=2$, however,
the emission of plasmons by quasiparticles destroys the
Fermi liquid fixed point.

In one dimension, even a short-range interaction drives the fermion
system to a different fixed point, which is a Luttinger liquid.
The formalism developed in Sec. \ref{sec:bose} can be used to diagonalize
the interacting fermion problem exactly and to access the Luttinger
liquid. As an example we consider an interaction which makes a
contribution to the action of the form
\begin{equation}
S_{\it int} = -{f_0\over 2}\int {d\omega \over 2\pi}
{}~{dq \over 2\pi}\sum_{S,T=L,R}
 J(S;q )~J(T;-q )
\end{equation}
where $L$ and $R$ specify the left and right Fermi points.
We introduce a fictitious field, $A_0(q)$, which generates this
interaction when integrated out of the partition function.
\begin{equation}
S^{\prime}_{\it int} = \int {d\omega \over 2\pi}\int {dq \over 2\pi}
\bigg{\{} {1\over 2f_0} A_0(q )~A_0(-q)
+ \big{[} J(L; q) + J(R; q)\big{]}~A_0(-q) \bigg{\}}~ .
\end{equation}
Now the total action has the form discussed in Sec. \ref{sec:bose} and
we follow the procedure given there to determine the boson correlation
function. We find the propagator for the right hand patch to
be given by
\begin{equation}
\langle a(R;q )~a^{\dag}(R;q)\rangle
= {i\over 2}\Big{\{} \Big{[}1+{1\over \sqrt{1-F_0^2/4}} \Big{]}
{1\over \omega -q v_F^{\prime} \sqrt{1-F_0^2/4}} +
\Big{[}1-{1\over \sqrt{1-F_0^2/4}} \Big{]}
{1\over \omega +q v_F^{\prime} \sqrt{1-F_0^2/4}} \Big{\}}
\end{equation}
where $v_F^{\prime} = v_F + f_0/(2\pi)$ is a renormalized Fermi velocity due
to intrapatch interactions and $F_0= f_0/(\pi v_F^{\prime})$. If we now
define $\eta $ such that
\begin{equation}
\cosh ^2 \eta = {1\over 2}\big{[} 1+ {1\over \sqrt{1-F_0^2/4}}
\big{]}
\end{equation}
the boson propagator can be written as
\begin{eqnarray}
\langle a(R;q )~a^{\dag}(R;q )\rangle
&=& {i\over 2}\Big{\{} \cosh ^2 \eta ~
{1\over \omega -q v_F^{\prime} \sqrt{1-F_0^2/4}} -
\sinh ^2 \eta ~
{1\over \omega +q v_F^{\prime} \sqrt{1-F_0^2/4}} \Big{\}}
\end{eqnarray}
where $\tanh 2\eta = F_0$. The fermion propagator found from
Eq. (\ref{Gff}) is non-analytic in the coupling constant $f_0$ and
cannot be obtained  by perturbing in the coupling constant,
\begin{equation}
G_F(R;x,t) \propto {{(x^2 + v_F^{*2} t^2)^{-\alpha}}\over{x - v_F^* t}}
\end{equation}
where the anomalous exponent $\alpha = \sinh ^2 \eta $ and the Fermi
velocity $v_F^*=v_F^{\prime}\sqrt{1-F_0^2/4}$.
The result is exact and identical
to that found by diagonalizing  the bosonic Hamiltonian
with a Bogoliubov transformation\cite{Tony}.

As a second example, we consider a fermion liquid in two dimensions
with short-range interactions. In earlier work\cite{HKM} we discussed,
by a different but equivalent approach, the effect of an interaction
which could be parameterized by a single Landau parameter, $f_0$.
It was shown that a perturbative expansion in
$f_0$ gave for the imaginary part of the self-energy a term proportional to
$f_0^2 \omega ^2 \ln |\omega |$ in two dimensions and $f_0^2 \omega ^2$ in
three
dimensions as in Fermi liquid theory.  A non-perturbative treatment of
this problem, involving a numerical computation of the spectral function,
is presented in Appendix A.  This calculation confirms that the
low-order perturbative result
is qualitatively correct, even at large $f_0$.  Also in this Appendix we
show how the Green's functions of the $f_0$ problem are obtained within the
present generating function approach.  Here, as a further illustration of
this approach, we consider in detail a Fermi liquid with a $f_1$
interaction.  The bosonized $f_1$ interaction contributes to the
action the following term:
\begin{equation}
S_{FL}=-{f_1\over 2Vk_F^2}\int{d\omega \over 2\pi }\sum_{S,T,\bf q}
J(S;q)~{\bf k}_S\cdot {\bf k}_T~J(T;-q)
\label{FL1}
\end{equation}
which can be separated into two parts, one longitudinal and the
other transverse:
\begin{eqnarray}
S_{FL}={f_1\over 2Vk_F^2}\int{d\omega \over 2\pi }\sum_{S,T,\bf q}
J(S;q)~\Big{\{ }{\bf {[k_{\it S}\cdot q]\over |q|}
{[k_{\it T}\cdot (-q)]\over |q|}+{[k_{\it S}\times q]\over |q|}
{[k_{\it T}\times (-q)]\over |q|}}\Big{\} } ~J(T;-q) ~ .
\label{FL2}
\end{eqnarray}
Introducing two fictitious fields $A_l$ and $A_t$ which generate the
interaction when integrated out of the partition function,
the total action is given by:
\begin{eqnarray}
S[A_{l,t},a] = \sum_{ S} \sum_{{\bf q, q\cdot \hat{n}}_S >0}
\int {d\omega \over 2\pi}
{}~(\omega - v_F^*~ {\bf q\cdot \hat{n}_{\it S}})~a^{*}( S; q)~a( S; q)
+S^{\prime }_{FL}
\end{eqnarray}
where
\begin{eqnarray}
S^{\prime }_{FL} &=&  -{1\over 2~f_1}\int{d\omega \over 2\pi }\int {d^2q\over
(2\pi )^2}~[A_{l}(q)~A_{l}(-q)+A_{t}(q)~A_{t}(-q)]  \nonumber \\
&+& {1\over k_F~V}\int {d\omega \over 2\pi }\sum_S \sum_{\bf q}\big{[}
{\bf k_{\it S}\cdot q \over |q|}~J(S;q)~A_{l}(-q) +
{\bf k_{\it S}\times q \over |q|}~J(S;q)~A_{t}(-q) \big{]} ~ .
\end{eqnarray}
As usual, we contruct the generating functional and integrate out
the boson fields, $a$ and $a^*$, to obtain the effective
action for the $A_{l,t}$ fields:
\begin{eqnarray}
S_{eff}[A_{l,t}] = {1\over 2}\int {d\omega \over 2\pi }\int {d^2q
\over (2\pi )^2}~\big{\{ } [-{1\over f_1}+\chi _l(q)]~A_l(q)~A_l(-q)
+[-{1\over f_1}+\chi _t(q)]~A_t(q)~A_t(-q) \big{\} } ~ ,
\end{eqnarray}
where
\begin{eqnarray}
\chi_l(q)
&=& N^*(0)\big{[}-x^2-{1\over 2}+\theta(x^2-1)|x|{x^2\over \sqrt{x^2-1} }
-i\theta(1-x^2)|x|{x^2\over \sqrt{1-x^2} }\big{]} ~ , \nonumber \\
\chi_t(q)
&=& N^*(0)\big{[}x^2-{1\over 2}-\theta(x^2-1)|x| \sqrt{x^2-1}
-i\theta(1-x^2)|x| \sqrt{1-x^2} \big{]} ~ .
\end{eqnarray}
The boson propagator is given by
\begin{eqnarray}
\langle a(S;q)~a^{\dag}(S;q)\rangle &=&
{i\over \omega - v_F^* {\bf q\cdot \hat{n}_{\it S}}+i\eta ~{\rm sgn}(\omega )}
\nonumber \\
&-& i{\Lambda \over (2\pi )^2 } {\bf q\cdot \hat{n}_{\it S}\over |q|^2}
{{({\bf q\cdot \hat{n}_{\it S}})^2 D_l(q)+
({\bf q\times \hat{n}_{\it S}})^2 D_t(q)}
\over{[\omega -v_F^* {\bf q\cdot \hat{n}_{\it S}}+i\eta ~{\rm sgn}(\omega
)]^2}}~ .
\end{eqnarray}
where the propagators of the mediating fields are given by
\begin{eqnarray}
D_l(q) &=& {1\over -1/f_1 + \chi ^l(q)} ~ ,  \\
D_t(q) &=& {1\over -1/f_1 + \chi ^t(q)} ~ .
\end{eqnarray}
These results will be used in the discussion of the physics of the
half-filled Landau level given in the next section and in Appendix C.

To complete this section we consider the effect of long-range
longitudinal interactions
such as the Coulomb interaction
or the super long-range interaction of Bares and Wen.\cite{Wen}
Following the procedure given in Sec. \ref{sec:bose}, we obtain the fermion
Green's function by setting $K^0_{00}=1/V(q)$ in Eq. (\ref{Kmn})
and eliminating the transverse gauge fields.
We briefly review what we find from this analysis and refer the
reader to a previous paper\cite{HKMS} for details.
The correction to the boson propagator is given by
\begin{eqnarray}
\delta G_B({\bf S; q}, \omega) &=&
i~ {{V({\bf q})}\over{
1 + V({\bf q}) \chi ^0(q)}}
{{1}\over{[\omega - v_F^* {\bf \hat{n}_S \cdot q} +i\eta ~{\rm
sgn}(\omega)]^2}}
\end{eqnarray}
and the effect of screening is manifest.
For example, in the case of the Coulomb interaction, $V({\bf q}) \propto
e^2 |{\bf q}|^{1-D}$ and consequently, in dimensions greater than one,
$\delta G_B$ is independent of the coupling constant $e^2$ in the limit of
small
$\bf q$ and small $\omega$.
We can calculate the correction to the fermion Green's function by
expanding in powers of $\delta G_B$. To first order,
we obtain the well-known
result that the imaginary part of the fermion self-energy is proportional to
$(\omega^2 / \epsilon_F)~ \ln |\omega|$ in $D=2$,
and $(\lambda / \epsilon_F k_F)~ \omega ^2$ in $D=3$,
in agreement with Fermi liquid theory\cite{HKM}.
Below we will show that in the case of Coulomb interactions
the weight of the quasiparticle pole, $Z_F$, remains non-zero at
the Fermi surface and the Landau Fermi liquid is stable
as expected.  It might be thought that this is the case for all types of
longitudinal interactions, as it appears to be a consequence of screening.
However, super long-range interactions {\it can} destroy the Fermi
liquid state.  Bares and Wen\cite{Wen} considered a system of fermions in
two dimensions interacting via a logarithmic potential,
$V({\bf q}) = g/{\bf q}^2$, and showed that
within the random phase approximation (RPA) $Z_F = 0$.
Note that $g$ is an energy scale and the plasmon gap is non-zero due to the
super-long-range nature of the interaction and is given by
$\omega_p =v_F \sqrt{m g/4\pi}$.  (By Kohn's theorem, the gap depends on
the {\it bare} fermion mass and Fermi velocity.  See Appendix C for details on
how Kohn's theorem is automatically recovered within multidimensional
bosonization.)

One might be tempted to compute $Z_F$ by employing the
Kramers-Kronig relation to derive the real part of the self-energy from the
imaginary part estimated here.  This procedure, however, is unreliable, as the
Kramers-Kronig relations involve an integral over all frequencies, whereas the
above calculation ignores high-energy processes such as the emission of
plasmons by quasiparticles.  Instead we compute the real-space, equal-time,
two-point fermion Green's function directly.  Because we wish to examine the
effects of plasmons, we take the energy cutoff to be much larger
than the plasmon energy, $v_F^* \lambda \gg \omega_p$.  Obviously, then,
bosonization is less accurate for systems with a large plasmon gap.
Nevertheless, we expect it to be qualitatively correct, as discussed below.
Setting $t = 0$, $x_\perp \equiv {\bf x \times \hat{n}}_S = 0$ and
introducing $x_\parallel \equiv {\bf x \cdot \hat{n}}_S$,
$q_\parallel \equiv {\bf q \cdot \hat{n}}_S$, and
$q_\perp \equiv {\bf q \times \hat{n}}_S$ we find:
\begin{eqnarray}
G_F({S; x_\parallel}) &\approx& {{e^{i k_F x_\parallel}}\over{x_\parallel}}
\exp \Big{\{} i~ \int_{-\lambda/2}^{\lambda/2} {{dq_\parallel}\over{2 \pi}}~
\int_{-\lambda/2}^{\lambda/2} {{dq_\perp}\over{2 \pi}}
{{\exp(i q_\parallel x_\parallel) - 1}\over{q^2}}~
\nonumber \\
&\times& \int^{\lambda}_{|{\bf q}|} {{d\omega}\over{\pi}}~
{{1}\over{1/g - v_F^{2} m /(4 \pi \omega^2)}}~
{{1} \over{[\omega - v_F^* q_\parallel + i \eta~ {\rm sgn}(\omega)]^2}}
\Big{\}}~ .
\end{eqnarray}
Here because plasmons are important only in the large-$x$ limit, we
have expanded the Lindhard function $\chi^0$ in powers of $1/x$,
$x \equiv \omega / (v_F^* |{\bf q}|)$:
\begin{equation}
\chi^0(x) = - {{N^*(0)}\over{2 x^2}} + O(1/x^4)\ + i\eta~
\end{equation}
Kohn's theorem is respected by replacing the effective mass
and velocity in the plasmon pole by bare quantities.  Now
since $\omega > v_F^* |{\bf q}|$, in the frequency integral we may make the
approximation
of replacing $[\omega - v_F^* q_\parallel + i \eta~ {\rm sgn}(\omega)]^2$
by $\omega^2 + i \eta^\prime$ and carry
out the integrations.  The integral over $\omega$ yields a constant due
to the plasmon pole at non-zero frequency.
Logarithmic dependence on $x_\parallel$
then is due to the factor of $q^2 = q_\perp^2 + q_\parallel^2$
in the denominator of the momentum integral.
Upon exponentiating the boson Green's
function we obtain the fermion quasiparticle propagator in configuration space:
\begin{eqnarray}
G_F( S; x_{\parallel}) &\approx & e^{i k_F x_\parallel}~
{(\lambda |x_{\parallel}|)^{-\zeta}
\over x_{\parallel}} ~ ; ~ \lambda |x_{\parallel}| \gg 1
\nonumber \\
&\approx & {{e^{i k_F x_\parallel}}\over{x_{\parallel}}}~ ;~
\lambda |x_{\parallel}| \ll 1
\end{eqnarray}
where
\begin{equation}
\zeta ={1\over 2}\sqrt{g \over \pi v_F k_F}~ .
\end{equation}
The non-analytic dependence on $x_{\parallel}$ demonstrates explicitly
the destruction of the quasiparticle pole by the emission of plasmons.
Power law behavior with exponent $\zeta $, rather
than the logarithmic behavior found in RPA, is a consequence of bosonization
which treats the interaction non-perturbatively.
In fact, Bares and Wen\cite{Wen} used both a
hydrodynamic calculation and a variational wavefunction to argue that the
logarithm is promoted to a power law, with exponent $\zeta /\sqrt{2}$,
the difference by a factor of $1/\sqrt{2}$ is due to spin degeneracy which
is not taken into account here.
This is further evidence that bosonization can accurately describe
non-Fermi liquid systems.
In the opposite limit of $\omega_p |t| \gg 1$, however,
an evaluation of the fermion propagator for a system of area $L^2$
yields the usual $(x_\parallel - v_F^* t)^{-1}$ Fermi liquid form,
multiplied by a prefactor of $(\lambda L)^{- \zeta}$.
As screening is effective at low frequencies,
thermodynamic properties such as the specific heat which
are defined in the equilibrium $q$-limit
also retain the usual Fermi liquid form.  Note that in either
limit the propagator is odd under the combined CPT operation of
$({\bf x}, t) \rightarrow -({\bf x}, t)$
and complex conjugation as demanded by Fermi statistics.
The anomalous exponent $\zeta$ in the equal-time two-point
function shows that the discontinuity in the quasiparticle occupancy at the
Fermi surface has been replaced by a continuous, though non-analytic,
crossover.
The quasiparticle occupancy decreases from 1 to 0 over a momentum
scale of order $\lambda$.  Since the discontinuity has been smeared out, we
might expect
to encounter problems of consistency as the derivation of the current algebra
Eq. (\ref{u1alge}) relies upon the
existence of an abrupt change in the quasiparticle occupancy over a
momentum scale much smaller than $\lambda$.  Nevertheless, as noted above,
the form of the Green's function agrees with that found by
an independent argument using a variational wavefunction\cite{Wen}.
The approximations implicit in bosonization (restriction to low energy
processes,
the existence of the discontinuity, and assumed stability of the fixed point)
are robust in this case.

It is instructive to repeat the calculation
of the equal time Green's function for the case of a Coulomb interaction in
both two and three dimensions.  In two dimensions plasmons are found at
$\omega \propto \sqrt{|{\bf q}|}$.  In this case the integral over $\omega$
is no longer constant, as it was for the Bares and Wen interaction, but
rather is proportional to $\sqrt{|{\bf q}|}$ and vanishes in the limit
of small momentum.  The momentum integral therefore
no longer diverges logarithmically at large-$x_\parallel$; rather
the integral is proportional to $\sqrt{\lambda/k_F} \rightarrow 0$.
While there is a non-zero gap for plasmon excitations in three dimensions,
the integral over $\bf q$ again is suppressed, in this case because of
the reduced phase space at small momentum and consequently
$\delta G_B \propto \lambda/k_F \rightarrow 0$.
In either case the quasiparticle propagator has the
standard Fermi liquid form.  The weight of the pole undergoes no additional
renormalization beyond the factor of $Z_{\bf k}$ due to integrating
out the high-energy degrees of freedom, Eq. (\ref{ZK}).
This is as expected, since the renormalization of the quasiparticle pole in
Landau Fermi liquids is due to high-energy processes, not
the low-energy processes found near the Fermi surface.

In this section we examined the effects of both short- and long-range
longitudinal interactions on fermion liquids.
We find that the Fermi liquid state is the only solution to the problem
of a degenerate gas of fermions interacting via short-range or Coulomb
two-body interaction in two and three spatial dimensions.
Bosonization allows us to go beyond an assumed Fermi
liquid form for the quasiparticle propagator.  Indeed, had we considered
fermions with spin, we would have found that
the bosonized Hamiltonian separates into charge and
spin parts\cite{HKM}, $H = H_c + H_s$, leading to the possibility that,
as in one dimension\cite{Tony}, the quasiparticle propagator would also
exhibit spin-charge separation.  Spin-charge separation in dimensions larger
than one would, however, destroy the Fermi liquid,
as the key element, the existence of a
pole in the single-particle Green's function with non-zero spectral weight,
would be replaced by a branch cut and the pole would be destroyed.
This does not occur because the location of the pole of the boson
propagator is unchanged from its
free value in the $\Lambda \rightarrow 0$ limit.  Consequently the spin and
charge velocities are equal and both degrees of freedom
propagate together in the usual quasiparticle form\cite{HKMS}.
On the other hand,
the Fermi liquid form is destroyed both in one dimension, where a Luttinger
liquid occurs instead, and in the case of the super long-range
interaction in two dimensions studied by Bares and Wen.  The non-Fermi liquid
fixed points of these two examples are accessible via bosonization.  We
therefore have reason to expect that bosonization can describe other
non-Fermi liquid states of matter.  The system we study in the next section
is another such example.

\section{Transverse Gauge Fields: the Half-Filled Landau Level}
\label{sec:transverse}

The two dimensional electron liquid in a magnetic field with even
denominator filling fraction has been the subject of extensive studies.
Experiments at even denominator fillings point to a novel state which
is compressible unlike the incompressible states found at
odd denominator fractional fillings.\cite{Jiang,Willett,Du,Leadley,Goldman}
Attempts have been made to understand the new quantum state by
constructing composite particles which are made of magnetic flux tubes
attached to each electron\cite{Kalmeyer,Lopez,Halperin}.
The composite particles see
zero average magnetic field and in many respects behave like real particles.
In this section we study the specific case of filling fraction $\nu =1/2$.

\subsection{Half-Filled Landau Level}
\label{subsec:hfll}

The Hamiltonian for a two-dimensional electron gas of density $\rho_f$
in a perpendicular magnetic field $B$ is given by:
\begin{equation}
H=\int d^2x~c_e^{\dag}({\bf x }){(-i{\bf \nabla +({\it e\over c})A})^2
\over 2m} c_e({\bf x}) + H_{Coulomb}~ .
\end{equation}
Here $B={\bf \nabla \times A}$;
at a value $B=4\pi \rho_f c / e$ half of the states in the
first Landau level are filled.
The HLR theory\cite{Halperin} makes use of a local gauge transformation
to describe this system as a collection of
quasiparticles which obey Fermi statistics:
\begin{equation}
c^{\dag}({\bf x})=c_e^{\dag}({\bf x})~\exp \Big{[}-i\tilde{\phi} \int
d^2y~{\rm arg}({\bf x-y})~c_e^{\dag}({\bf y})~c_e({\bf y})\Big{]} ~ .
\label{quasi}
\end{equation}
When $\tilde{\phi}$ is an even integer the quasiparticles are fermions.
Each quasiparticle is a composite object consisting of the physical electron
together with a flux tube of integer $\tilde{\phi}$ quanta.
The transformed Hamiltonian is given by
\begin{equation}
H=\int d^2x~c^{\dag}({\bf x }){(-i{\bf \nabla +({\it e\over c})A-A^{\prime}
})^2 \over 2m} c({\bf x}) ~ ,
\end{equation}
where
\begin{equation}
{\bf A^{\prime}(x)}=\tilde{\phi}\int d^2y {\bf \hat{z} \times (x-y)\over
(x-y)^{\rm 2}}~c^{\dag}({\bf y})~c({\bf y}) ~ .
\end{equation}
The average field strength of the flux tubes can now be adjusted to
cancel out the external magnetic field.  At half-filling setting
$\tilde{\phi}=2$
\begin{equation}
{\bf \nabla \times \Big{[}({\it e\over c})A-A^{\prime}\Big{]}}
={e\over c}\Big{(}B-{2\pi \rho_f c\tilde{\phi}\over e}\Big{)}=0  ~ ,
\end{equation}
consequently, at the mean-field level, the quasiparticles behave as if
they are free fermions in
zero net field; the infinite Landau degeneracy is lifted and
presumably the state is stable against two-body Coulomb interactions.
Now the important question is whether gauge fluctuations (or equivalently
quasiparticle density fluctuations) modify
or destroy this mean-field state.  Unlike physical Maxwell electromagnetic
gauge fields for which the gauge coupling is the fine structure
constant $\alpha \approx 1/137$, the coupling constant between the composite
fermions and
the transverse component of the statistical gauge field is of order unity.

In the following we assume that despite strong gauge fluctuations
the bosonic construction continues
to apply.  Later we will verify that this assumption is
correct for the case of a Coulomb interaction.
We determine the composite quasiparticle Green's function
and show that the physics of the $\nu = 1/2$ state is controlled by
a marginal Fermi liquid fixed point.
It should be noted that the gauge fields of the HLR theory are compact
and consequently instantons may play a role in the determination
of the physical {\it electron} Green's function\cite{Kim}.
The action of this system is given by\cite{Halperin}:
\begin{eqnarray}
S &=& \int d^2x~dt~\big{[} c^{*}(i~\partial_t +A_0)c
+ {1\over 2m} c^{*}({\bf \nabla }+i~({e\over c})
{\bf A}-i~{\bf A^{\prime}})^2c
+ {A_0\over 2\pi \tilde{\phi}} ({\bf \nabla \times A^{\prime}})
\big{]} \nonumber \\
&-& {1\over 2}~{1\over (2\pi \tilde{\phi })^2}\int d^2x~d^2y~dt~
{\bf [\nabla_x \times A^{\prime}(x)]
}~V({\bf x -y})~{\bf [\nabla_y \times A^{\prime}(y)]}~ ,
\end{eqnarray}
The last term in the action
is the longitudinal two-body Coulomb interaction which now is written in
terms of the gauge field by making use of the constraint that a
flux tube is attached
to each electron\cite{Halperin}, ${\bf \nabla \times A^{\prime}}
=-2\pi \tilde{\phi}~ c^{*}~c$.  It is important to notice that
the longitudinal interaction plays a crucial role in the Chern-Simons gauge
theory for it suppresses fermion density fluctuations.  Since density
fluctuations are equivalent to gauge fluctuations in a Chern-Simons theory,
repulsive long-range longitudinal interactions moderate and control the gauge
interactions.
After integrating out the high energy
quasiparticles following the procedure of Sec. \ref{sec:bose},
and adopting a change of notation
${\bf A^{\prime}} - (e / c){\bf A} \rightarrow {\bf A}$,
the HLR effective action reads:
\begin{eqnarray}
S &=& \int d^2x~dt~\big{[} \psi ^{*}(i~\partial_t +A_0)\psi
+ {1\over 2m^*}\psi ^{*}({\bf \nabla }-i~{\bf A})^2\psi
+ {A_0\over 2\pi \tilde{\phi}} ({\bf \nabla \times A}) \big{]} \nonumber \\
&-& {1\over 2}~{1\over (2\pi \tilde{\phi })^2}\int d^2x~d^2y~dt~
{\bf [\nabla_x \times A(x)]}~V({\bf x -y})~{\bf [\nabla_y \times A(y)] }~ .
\label{SFQH}
\end{eqnarray}
At filling fraction $\nu =1/2$ we set $\tilde{\phi} =2$ to cancel out the
external field; now ${\bf A}$ should be interpreted as the fluctuation
about this mean field value.
Fermi liquid interactions should be included in the action also.
If all coefficients except $f_1$ are set equal to zero, they are described
by the term
\begin{equation}
S_{FL} = -{{f_1}\over{2 V k_F^2}}~ \int {{d\omega}\over{2\pi}}~
\sum_{S,T,{\bf q}}~ J(S; q)~ {\bf k_{\it S} \cdot k_{\it T}}~ J(T; -q)
\label{f1}
\end{equation}
which is made gauge covariant by the replacement\cite{Simon}
${\bf k} = -i{\bf \nabla }\rightarrow -i {\bf \nabla - A}$.
The bare gauge action, Eq.(\ref{SG0})
can now be read off from Eq. (\ref{SFQH})
\begin{eqnarray}
S_G^0 &=& {i\over 4 \pi \tilde{\phi}}
\int {d^2q\over (2\pi)^2}\int {d\omega \over 2\pi }~
\{ A_0(-q)~ [{\bf q \times A}(q)] - A_0(q)~ [{\bf q \times A}(-q)] \}
\nonumber \\
&-& {1\over 2}~{1\over (2\pi \tilde{\phi})^2} \int {d^2q\over (2\pi)^2}
\int {d\omega \over 2\pi }~{\bf q}^2~ V({\bf q})~
[{\bf A}(q)~{\bf \cdot A}(-q)]~ ,
\label{CS}
\end{eqnarray}
here $V({\bf q})$ is the Fourier transform of $V({\bf x -y})$; for the
Coulomb interaction, $V({\bf q}) = 2\pi e^2 / |{\bf q}|$.
For computational convenience, we work in the Coulomb gauge which
eliminates the longitudinal component of the vector gauge field, only
the transverse component $A^T(q) = {\bf q\times A}(q)/|{\bf q}|$ remains.

\subsection{Bosonization of Half-Filled Landau Level}
\label{subsec:hfllbose}

We now use the bosonization method to
determine the composite quasiparticle two-point Green's function.
As before, we bosonize the action
Eq. (\ref{SFQH}), construct the generating functional $Z[\xi,\xi^*]$,
and integrate out the boson fields to calculate the effective gauge action.
The components of the inverse
gauge propagator $K_{\mu \nu }=[D^{-1}]_{\mu \nu }$ (now the indices
$\mu $ and $\nu $ run over only the time and transverse directions, $0$ and
$T$) are given by:
\begin{eqnarray}
K_{00}(q)&=&\chi ^0(q) \nonumber \\
K_{0T}(q)&=&{i\over 2\pi \tilde{\phi}}|{\bf q}| =K_{T0}(q) \nonumber \\
K_{TT}(q)&=&{{\bf q}^2\over (2\pi \tilde{\phi} )^2}V({\bf q})
+\chi ^T(q) ~ .
\label{Kmatrix}
\end{eqnarray}
Again the additive correction to the boson propagator is given by
Eq. (\ref{additive}) where $\epsilon ^{\mu}(  S;q)$
is now the two-component vector $(1,{\bf {q \times v^*_{\it S} / |q|}})$.
In the q-limit, $\omega \ll v^*_F|{\bf q}|$, the gauge propagator
is dominated by the transverse components and
\begin{equation}
D_{\mu \nu }(q)~\epsilon^\mu ( S;q) ~\epsilon^\nu ( S;-q)
\approx {{|{\bf q\times \hat{n}_{\it S}}|^2}\over{{\bf q}^2}}
{{v_F^{* 2}~ N^*(0)}\over{i \gamma N^*(0){ |\omega|\over |{\bf q}|}
- {{{\bf q}^2}\over{(2\pi \tilde{\phi})^2}} [1 + N^*(0)~ V({\bf q})]}}~ ;
\label{damp}
\end{equation}
here $\gamma = {\rho _f / \pi v_F^* N^*(0)}$ following the notation of
HLR\cite{Halperin}. In this limit the gauge propagator has an imaginary part
which is a result of quasiparticle damping.  For the Coulomb interaction,
\begin{equation}
D_{\mu \nu }(q)~\epsilon^\mu ( S;q) ~ \epsilon^\nu (S;-q)
\approx v_F^{*2}
{{|{\bf q\times \hat{n}_{\it S}}|^2}\over{{\bf q}^2}}
{1\over i\gamma {|\omega|\over |{\bf q}|}-\chi |{\bf q}|}
\end{equation}
where $\chi = {2\pi e^2 / (4 \pi^2 \epsilon {\tilde{\phi}}^2)}$,
$\epsilon$ is the dielectric constant,
and the pole is located at $\omega \approx i~e^2 {\bf q}^2/k_F$.
If, on the other hand, $V({\bf q})$ is a short-range interaction,
the pole is at $\omega \approx i~v_F^*~ |{\bf q}|^3/k_F^2$.
The pole location agrees
with the RPA result\cite{Halperin,Lee}.  In the opposite $\omega$-limit,
$\omega \gg v^*_F|{\bf q}|$, we find
\begin{equation}
D_{\mu \nu }(q)~\epsilon^\mu (S;q) ~\epsilon^\nu (S;-q)
\approx {1\over 2}~
{\omega ^2\over {\bf q}^2}~ {{v_F^{* 2}~ N^*(0)~ (2\pi \tilde{\phi})^2}\over
{\omega ^2 -  (2 \pi \tilde{\phi}~ \rho_f)^2/m^{* 2}}} ~ ;
\label{gap}
\end{equation}
in this limit the poles which characterize collective excitations\cite{HKM},
the cyclotron modes of the composite fermions,
are found at $|\omega | = {2\pi \rho_f \tilde{\phi} / m^*}$,
the cyclotron frequency of a free particle of mass $m^*$
in an external magnetic field of $B = {2\pi \rho_f c \tilde{\phi} / e}$.
As it stands this result violates Kohn's
theorem\cite{Kohn}.  The correct result is obtained when the Fermi
liquid interaction, Eq. (\ref{f1}), is included in the bosonized theory.
The poles are then shifted to $|\omega| = \omega_c = 2\pi \rho_f
({1 / m^*}+{f_1 / 4\pi }) \tilde{\phi}
= {2\pi \rho_f \tilde{\phi} / m_b}$ in accord with Kohn's theorem.
The details of the proof of this result are given in Appendix C.
The form of Eq.(\ref{gap}) is similar to that found for the
super long range interaction of Bares and Wen\cite{Wen} in the previous
section.  Since a projection onto the lowest Landau level is implicit in
the HLR theory, however, we must assume
$v_F^* \lambda \ll \omega_c$ so that
inter-Landau level transitions are not allowed.  This restriction is an
auxiliary condition which must be taken into account when defining
the composite fermion Green's function.  Non-Fermi liquid
behavior, if it exists, must have its origin in the behavior of the
gauge propagator in the opposite $q$-limit.
We may estimate the fermion quasiparticle self-energy\cite{HKM,HKMS}
from Eq. (\ref{damp}) by expanding the fermion Green's function
in powers of $D_{\mu \nu}$; however it should be noted that there is no obvious
small parameter to justify the expansion.  Given this caveat,
in the case of the Coulomb interaction at first order in the expansion we
obtain
the marginal Fermi liquid (MFL) form for the self-energy in agreement
with the RPA calculation of HLR\cite{Halperin}.

As the expansion in powers of the boson propagator is unreliable, it
is more useful to determine the quasiparticle propagator directly in
real space and time. We consider the fermion Green's function,
Eq.(\ref{Gffc}) and focus on the $q$-limit.
We now set $x_\perp = 0$ and factorize the quasiparticle Green's function
into the product of the free term
\begin{equation}
G^0_F( S;{\bf x},t)={e^{i~{\bf k_{\it S}\cdot x}}\over
{\bf x\cdot \hat{n}_{\it S}}-v^*_F~t}
\end{equation}
and a factor which is the exponential of the additive correction to
the boson propagator, $\delta G_B({\bf {\it S};x},t)$, Eq. (\ref{Gb}).
Evaluation of the integral over $q_{\parallel}$ yields
\begin{eqnarray}
\delta G_B({\bf {\it S};x},t) &=&
2 i |x_{\parallel}|{\zeta \over v^*_F}\int^{v_F^* \lambda / 2}_0
d\omega ~e^{i\omega~{\rm sgn}(x_{\parallel})(x_{\parallel}/v^*_F -t)}
\int ^{\lambda \over 2}_{\omega \over v^*_F}
{{q_{\perp}~ dq_\perp}\over{q^2_{\perp} - i |\omega|{\gamma \over \chi}}}
\nonumber \\
&\approx& i |x_{\parallel}| {\zeta \over v^*_F}\int ^{\infty}_0 d\omega~
e^{i\omega~ {\rm sgn}(x_{\parallel})(x_{\parallel}/v_F^* - t) -
2 \omega/(\lambda v_F^*)}~
\ln \bigg{(} {i v^*_F \delta +\omega \over \omega} \bigg{)} ~ ;
\label{int2D}
\end{eqnarray}
here
\begin{equation}
\delta \equiv {{e^2}\over{4 \epsilon~ v_F^* ~ {\tilde{\phi}}^2}}~
{{\lambda^2}\over{k_F}}
\label{delta}
\end{equation}
is a new momentum scale and
\begin{equation}
\zeta \equiv {{{\tilde{\phi}}^2 \epsilon~ v_F^*}\over{2\pi e^2}}
\label{zeta}
\end{equation}
is a dimensionless number.
The final integration over frequency can be carried out using the
integration formula
\begin{equation}
\int ^{\infty}_0 ~e^{-\mu x}\ln(\beta +x)~dx = {1\over \mu}[\ln \beta -
e^{\mu \beta}{\rm Ei}(-\mu \beta)]
\end{equation}
where ${\rm Ei}(x)$ is the exponential integral function. The result is
\begin{eqnarray}
\delta G_B({\bf {\it S};x},t)
&=& -{x_{\parallel} \zeta \over x_{\parallel}-v^*_F t+2ia~{\rm sgn}
(x_{\parallel})}\Big{\{} {1\over2}~ \ln \big{[}
(x_{\parallel}-v^*_F t)^2 \delta^2 + 4 a^2 \delta^2 \big{]} +\gamma
\nonumber \\
&-& e^{(x_{\parallel}-v^*_F t)\delta ~{\rm sgn}(x_{\parallel})
+2ia \delta}~ {\rm Ei}\Big{(}-(x_{\parallel}-v^*_F t)\delta ~
{\rm sgn}(x_{\parallel})-2ia \delta\Big{)} \Big{\}}
\label{dgbexact}
\end{eqnarray}
which, upon making use of the asymptotic expansion of the exponential integral
function, reduces to
\begin{eqnarray}
\delta G_B({\bf {\it S};x},t)
&\approx& \zeta | x_{\parallel}\delta |~\ln |x_{\parallel}-v^*_F t| \delta~ ;~
|x_{\parallel}-v^*_F t|\delta \ll 1
\nonumber \\
&\approx& -\zeta {x_{\parallel}\over x_{\parallel}-v^*_F t+2ia~{\rm sgn}
(x_{\parallel})}\Big{[} \gamma + \ln \big{|} x_{\parallel}-v^*_F t
\big{|} \delta \Big{]}~
;~|x_{\parallel}-v^*_F t| \delta \gg 1~ .
\label{deltagb}
\end{eqnarray}
Here $\gamma \approx 0.577216$ is Euler's constant and $a \equiv 1/\lambda$.

For simplicity, first consider the equal-time quasiparticle
propagator given at long distances
$|{\bf \hat{n}_{\it S}\cdot x}|\gg 1/\delta $ by:
\begin{equation}
G_F({\bf {\it S};x})
\approx {e^{i{\bf k_{\it S}\cdot x}}\over{\bf \hat{n}_{\it S}\cdot x}}~
(\delta ~ |{\bf \hat{n}_{\it S}\cdot x}|)^{-\zeta}~ ,
\end{equation}
and at short distances $|{\bf \hat{n}_{\it S}\cdot x}|\ll
1/\delta $ by the usual Fermi liquid form:
\begin{equation}
G_F({\bf {\it S};x})
\approx {e^{i{\bf k_{\it S}\cdot x}}\over{\bf \hat{n}_{\it S}\cdot x}}~ .
\end{equation}
At momentum scales less than $\delta$ the quasiparticle pole
is destroyed by transverse gauge interactions suggesting that
the $\nu = 1/2$ system is controlled by a novel non-Fermi liquid
fixed point.  Despite the appearance of the anomalous exponent $\zeta$, the
composite quasiparticle Green's function differs in an important way from the
Green's function found for the super long-range longitudinal
interaction in the previous
section.  The composite quasiparticle occupancy drops abruptly from one to
zero at the Fermi surface over a momentum scale $\delta \propto \lambda^2$
which is much less than the cutoff $\lambda $ as $\lambda \rightarrow 0$.
As the basic consistency condition is satisfied in this limit, bosonization
can provide a systematic approach to the study of the new fixed point.

For typical GaAs/AlGaAs samples
the carrier density $\rho_f$ is of order $10^{11} /cm^2$, the electron band
mass
$m_b = 0.068 m_e$, and the dielectric constant is given by $\epsilon = 12.8$.
However, the effective mass of the composite fermion quasiparticle has to
be estimated.  Since electron states in the lowest Landau level are
degenerate, the electron band mass does not play an important role.
Instead, the energy scale is determined by the Coulomb energy at
the magnetic length scale.  From dimensional analysis\cite{Halperin}
the effective mass of the quasiparticle must be therefore:
\begin{equation}
m^*= {\epsilon \over e^2}{\sqrt{4\pi \rho_f}\over C}
\end{equation}
where $C$ is a dimensionless constant to be determined either
numerically, or from experiment.
Experiments at $\nu = 1/2$ suggest\cite{Du,Leadley}
that $m^* \approx 10 m_b$ and hence $C \approx 0.10$.
For values in this range we find $\zeta = C {\tilde{\phi}}^2 / 2 \pi
\approx 0.05$.  The small value of the exponent $\zeta$ is consistent with
the abrupt change in the occupancy inferred from the experimental observation
of sharp cyclotron resonances\cite{Willett,Goldman}.
For the purpose of estimating $\delta$, we may take the cutoff to be of order
the Fermi momentum $\lambda \approx k_F$ and find
$\delta = \lambda^2/(4 C {\tilde{\phi}}^2 k_F) = O(k_F)$ as expected since
$k_F$ is the only physical momentum scale in the problem.  Note that the
Fermi energy, $\epsilon_F = k_F^2/(2 m^*)$, is still much smaller than
the cyclotron energy, $\omega_c = k_F^2 \tilde{\phi} / (2 m_b)$, and
hence the requirement $v_F^* \lambda << \omega_c$ is respected even for
$\lambda \approx k_F$.
In the opposite limit $|{\bf x}| = 0$ the quasiparticle Green's function
is given by the free form $G_F({\bf 0}, t) = 1/(v_F^*~ t)$.  Consequently
the density of states is the same as that of non-interacting fermions with
Fermi velocity $v_F^*$.

We examine briefly longitudinal interactions other than Coulomb.
Because longitudinal interactions control fermion density fluctuations
and hence, by the constraint, gauge fluctuations, we expect repulsive
interactions which are longer-ranged than Coulomb to stabilize the
Fermi liquid fixed point.
For a general interaction $V({\bf q}) \propto 1/|{\bf q}|^{1-y}$
this question has been addressed by Nayak and Wilczek\cite{Nayak}
using a renormalization group expansion for small $y$.
They concluded that for $y$ positive
(negative) the gauge interaction is relevant (irrelevant).  The case $y=0$
is marginal and they found logarithmic corrections to Fermi liquid
behavior.  Here we set $V({\bf q})={g |{\bf q}|^{y-1}}$ and repeat
the calculation of the equal-time Green's function.  We obtain:
\begin{eqnarray}
\delta  G_B({\bf {\it S};x}) \approx
i\int {dq_{\perp}\over 2\pi }\int_0^{\lambda /2} {d\omega \over 2\pi }
{}~({x_{\parallel}\over v_F^*})~e^{i \omega x_{\parallel} / v_F^*}
{}~{v_F^* |q_{\perp }|\over \gamma ~\omega
+i\chi ~|q_{\perp}|^{2+y}} ~ .
\end{eqnarray}
where $\chi \equiv {g / (2\pi \tilde{\phi})^2}$.
When $y \not = 0$, the correction to the boson propagator scales as
$\delta G_B \propto (|x_{\parallel}| \delta^\prime)^{y/(2 + y)}$
where
$\delta^{\prime} \equiv (\chi / v_F^* \gamma) (v_F^* / \chi)^{(2 + y)/y}$.
Therefore, for  $y<0$, the gauge interaction is irrelevant and
$\delta G_B \rightarrow 0$ as
$|x_{\parallel}| \rightarrow \infty$, and the Landau fixed point is
recovered.  For $y > 0$, however, we expect pronounced deviations
from the Landau fixed point when $|\delta G_B| \gg 1$, or equivalently
when $|x_{\parallel}|\gg 1/\delta^{\prime}$.
The small momentum region is controlled by a non-Fermi
liquid fixed point about which bosonization can tell us little as the
deviations
from Fermi liquid behavior are too large.  For example, the discontinuity
in the quasiparticle occupancy at the Fermi surface is completely
eliminated\cite{Khvesh} and inconsistencies develop within the formalism.
The problem of a pure gauge interaction in two dimensions,
considered by Lee and Nagaosa as a model for the cuprate
superconductors\cite{Lee}, is the special case $y = 1$.
Again this system is outside the scope of bosonization.

\subsection{A Marginal Fermi Liquid}
\label{subsec:mfl}

The real-space and time
composite fermion Green's function given in the preceding section
by Eqs. (\ref{fermi}, \ref{deltagb}) is the starting point for the
more detailed discussion of single-particle properties which we give
in this section.  In particular we show that the Green's function is of
a marginal Fermi liquid (MFL) type.  In this section for simplicity we
set $v_F^* = 1$ and denote $x_\parallel$ by $x$.  We first rewrite the
Green's function in the two
limits $|x - t| \delta \gg 1$ and $|x - t| \delta \ll 1$:
\begin{eqnarray}
G_F(x, t) &=& {{1}\over{x - t + i a~ {\rm sgn}(x)}}~ \exp \big{\{} \zeta
|x| \delta \ln |x - t| \delta \big{\}};\ |x - t| \delta \ll 1
\nonumber \\
&=& {{1}\over{x - t + i a~ {\rm sgn}(x)}}~ \exp \big{\{} - \zeta
{{x \ln |x - t| \delta}\over{x - t}}\big{\}};\  |x - t| \delta \gg 1.
\label{gxtmfl1}
\end{eqnarray}
Ideally we would Fourier transform the Green's function directly into
momentum and frequency space to extract the self-energy.
Unfortunately this task is quite difficult, so instead we
begin with the observation that the pole of the free fermion
Green's function at $x = t$ has been destroyed by the transverse interactions.
As $x \rightarrow t$, $G_F(x, t) \rightarrow
{{|x - t|^{\zeta \delta |x|}}\over{x - t}}$ so the pole at $x = t$ has been
eliminated.  To ascertain where the spectral weight is concentrated now,
we expand the Green's function in the large-time limit
$|t| \gg |x + \zeta x \ln(|x - t| \delta)|$
and show later that corrections arising from contributions at shorter times
$|t| \ll |x + \zeta x \ln(|x - t| \delta)|$ contribute to the incoherent
spectral weight.  Since in the large-time limit $|x/(x - t)| \ll 1$
we may Taylor-series expand the
exponential appearing in the second line of Eq. (\ref{gxtmfl1}).
Keeping the leading terms we obtain:
\begin{eqnarray}
G_F(x, t) &=& \big{(} {{1}\over{x - t + i a~ {\rm sgn}(x)}} \big{)}~
{{1}\over{1 + {{\zeta x}\over{x - t}}~ \ln (|x - t| \delta) + HOT}}
\nonumber \\
&=& {{1}\over{(x - t) + \zeta x \ln (|x - t| \delta) + i a~ {\rm sgn}(x) +
HOT}}
\label{gxtmfl2}
\end{eqnarray}
where $HOT$ denotes higher-order terms which are negligible in the $|t| \gg
|x|$ limit.
We may now Fourier transform this leading term to extract the self-energy.
Upon making the approximation that $\ln (|x - t| \delta) \approx \ln (|x|
\delta)$ near the pole in Eq. (\ref{gxtmfl2}), the integral over time
can be done:
\begin{eqnarray}
G_F(k, \omega) &=& \int dx dt {{e^{i(\omega t - k x)}}\over
{(x - t) + \zeta x \ln (|x - t| \delta) + i a~ {\rm sgn}(x)}}
\nonumber \\
&\approx& 2 \pi i~ \int dx [\theta(-x) \theta(\omega)
- \theta(x) \theta(-\omega)]~
e^{i(\omega - k) x}~ \exp[i \zeta \omega x \ln (|x| \delta)] \ .
\nonumber \\
&\approx& 2 \pi i~ \int dx
[\theta(-x) \theta(\omega) - \theta(x) \theta(-\omega)]~
e^{i(\omega - k) x}~  \sum_{n=0}^{\infty}~
{{[i \zeta \omega x \ln (|x| \delta)]^n}\over{n!}}\ .
\label{gkwmfl1}
\end{eqnarray}
Each term resulting from the Taylor series expansion of the exponential
can be integrated separately.  It is easy to show that:
\begin{equation}
\int_0^\infty dx~ e^{(i k - \eta) x} [x \ln (x \delta)]^n
= {{-1}\over{(i k - \eta)^{n+1}}}~
\big{\{}n! \ln^n [(\eta -i k)/\delta]
+ O(\ln^{n-1}[(\eta - i k)/\delta]) \big{\}}
\end{equation}
where $\eta > 0$ is a small term which ensures convergence.
Up to subleading corrections, the Green's function is:
\begin{eqnarray}
G_F(k, \omega) &=& {{2 \pi}\over{\omega - k}} \sum_{n=0}^{\infty}
\bigg{[} {{\zeta \omega}\over{\omega - k}}~ \ln [i(\omega - k)~
{\rm sgn}(\omega) /\delta] \bigg{]}^n
\nonumber \\
&=& {{2 \pi}\over{(\omega - k) - \zeta \omega
\ln [i (\omega - k)~ {\rm sgn}({\omega}) / \delta)]}}\ .
\label{gkwmfl2}
\end{eqnarray}
Hence the leading contribution to the self-energy of the quasiparticles is:
\begin{equation}
\Sigma(k, \omega) = \zeta \omega \ln (|\omega - k| / \delta)
+ {{i \pi}\over{2}} \zeta \omega~ {\rm sgn}[(\omega - k)~ \omega]\ .
\end{equation}
At the pole, $k \approx \omega - \zeta \omega \ln (|\omega |/\delta )$,
$|\omega| \ll |k|$, and to logarithmic accuracy,
the self-energy is given by
\begin{equation}
\Sigma(k, \omega) = \zeta \omega \ln (|\omega| / v_F^* \delta)
- {{i \pi}\over{2}} \zeta ~\omega \ ~ ,
\label{mflsigma}
\end{equation}
the marginal Fermi liquid form.
In Eq. (\ref{mflsigma}) we have restored $v_F^*$.  Note that
$v_F^* = k_F/m^*$ is the Fermi velocity obtained earlier by integrating out the
high-energy degrees of freedom.  In a marginal Fermi liquid such as this,
there is no definite velocity; rather the velocity is a logarithmic function
of the energy scale and is of order, but not equal to, $v_F^*$.
At half-filling, $\tilde{\phi} = 2$ and $\zeta = 2 v_F^* / \pi e^2$.
Upon setting $\lambda \approx k_F$ for the purpose of estimating the size of
the
self-energy we find $v_F^* \delta = O(\epsilon_F)$, the
Fermi energy; hence Eq. (\ref{mflsigma})
is identical to that derived by Ioffe {\it et al.} by a
1/N perturbation expansion\cite{IKL}.

The evaluation of the self-energy might be questioned, in
particular because the time integral in Eq. (\ref{gkwmfl1})
extends over the whole
axis, whereas the approximate Green's function is accurate only at large
time.  Indeed, $G_F(x, t)$ does not have a pole at
$t = x + \zeta x \ln(|x - t| \delta)$ as can be seen from inspection of
Eq. (\ref{dgbexact}).
To see why nevertheless the calculation is qualitatively correct, consider
what happens if we restrict the time integral to large times
by excluding intermediate times $0 < t < 2 t_p$ for which the expansion of
Eq. (\ref{gxtmfl2}) is unreliable.  Here
$t_p \equiv x + \zeta x \ln(|x| \delta)$ and we can assume $x > 0$
without loss of generality.  The restricted
integral over time may be evaluated giving:
\begin{eqnarray}
\int^0_{-\infty} {{e^{-i \omega t}~ dt}\over{t - t_p - i a}}
+ \int^\infty_{2t_p} {{e^{-i \omega t}~ dt}\over{t - t_p - i a}}
&=& e^{(a - i t_p) \omega}~ \big{\{} 2 \pi i \theta(-\omega)
+ {\rm Ei}[(i t_p - a) \omega] - {\rm Ei}[-(i t_p + a) \omega] \big{\}}~
\nonumber \\
&=& e^{(a - i t_p) \omega}~ \big{\{} 2 \pi i \theta(-\omega)
- \pi i + 2 i \omega t_p + O(i \omega^3 t_p^3) \big{\}}
\end{eqnarray}
where the second line follows from an expansion of the exponential integral
about the origin
${\rm Ei}(z) = \gamma + \ln(z) + z + z^2/4 + O(z^3)$ with the assumption that
$t_p \gg a$.  By neglecting the term proportional to $2 i \omega t_p$
and higher order terms which lead only to subleading corrections,
the integral over $x$ can be done.  The resulting
spectral weight is identical to
that found from the imaginary part of Eq. (\ref{gkwmfl2}). Apparently, coherent
spectral weight in the $(k, \omega)$ plane is found in the long-time tail
of the real-space and time MFL Green's function despite the absence of
a pole in the $(x, t)$ plane.

In recent work Altshuler
{\it et al.} \cite{Altshuler} have determined the quasiparticle propagator
by a $1/N$ perturbation theory in the fermion basis.  In this
work it is claimed
that, because nonlinear terms in the fermion dispersion due to
curvature of the Fermi surface are neglected in the
bosonization scheme, the scheme cannot be used with confidence to
study a transverse gauge theory.
On the contrary, in the problem of the half-filled Landau level with
a Coulomb interaction, the self-energy Eq. (\ref{mflsigma}) is identical
to that found by Altshuler {\it et al.} \cite{Altshuler}
Indeed, the only difference between the two calculations
is the anomalous power law decay of the equal-time
propagator discussed in the previous section which has its origin in the
incoherent part of the quasiparticle spectrum and therefore should not be
expected to appear in analyses such as that of Altshuler {\it et al.}
which consider only the low frequency behavior of the Green's function.
In the case of Coulomb interactions ($y = 0$) nonlinear terms in
the fermion dispersion due to Fermi surface curvature
do not affect the low-energy quasiparticle spectrum in any significant way.
Why this should be the case is examined in the next subsection.

\subsection{Consistency of Bosonization: Nonlinear Dispersion and Vertex
Corrections}
\label{subsec:ward}

An important feature of the bosonization procedure is the linearization
of the free fermion spectrum in the vicinity of the Fermi surface:
$\epsilon_{\bf k_S+q}-\epsilon_{\bf k_S} \approx {\bf v_S^* \cdot q}$.
Nonlinear terms in the dispersion due to curvature of the Fermi surface
are neglected
and within bosonization can only be accounted for perturbatively\cite{old}.
This approximation might be questioned, particularly in the case of
a mediating transverse gauge interaction, when the interaction
vertex draws its most important contribution from $q_{\perp}$, the
momentum perpendicular to ${\bf k_S}$. To see this, consider the
generic gauge propagator $D^{-1} = i\gamma |\nu| / |{\bf q}| -
\chi |{\bf q}|^{1+y}$.  The quadratic curvature term scales as
$q_{\perp}^2 \propto \nu^{2\over 2+y}$ and therefore it, rather than
the pole familiar from the linearized approximation, $v_F^* q_\parallel$,
might dominate the physics for $y \ge 0$.
This question is best addressed in the fermion basis by
the consideration of the Ward Identities derived by
Castellani, {\it et al.}\cite{Castellani}. These authors found that
the Green's function derived using the Ward Identities is identical
to that found by bosonization provided that the relationship
between density and current vertices,
${\bf \Lambda}_S={\bf v_S^*}\Lambda _S^0$,
which becomes exact in one dimension, holds approximately in
dimensions greater than one. In the same context, we may
ask also whether deviations from this relationship
are so singular that they destroy the form of the
propagator found both within bosonization and the Ward Identity approach.
For the Chern-Simons gauge theory with $y=0$, the physically
relevant case of the long-range Coulomb interaction, we show that the
bosonization scheme is internally self-consistent, and that the results
derived earlier in this section continue to hold.

We start by deriving the Ward Identities which relate the density and current
vertices of the bosonized interacting fermion system in two dimensions;
for simplicity,
we give the proof for a fermion liquid interacting via a transverse
gauge interaction only. The result is readily shown to be true in
general.
We introduce the time-ordered current-current correlation function
between currents in different patches
\begin{equation}
\Pi_{S;Q}({\bf q})=-{i\over V}\langle {\cal{T}}~J({\bf {\it S};q})
{}~J({\bf {\it Q};-q})
\rangle
\end{equation}
and the amputated vertex function
\begin{equation}
\Gamma _{S;T}({\bf p_1,p_2;q})=-\langle {\cal{T}}~J({\bf {\it S};-q})
{}~\psi({\bf {\it T};p_1})~\psi^{\dag}({\bf {\it T};p_2})\rangle_{\rm amp}
\end{equation}
where ${\bf q+p_1-p_2=0}$. Then using the current algebra,
we derive their equations of motion:
\begin{eqnarray}
(\nu -{\bf q\cdot v^*}_S)\Gamma _{S;T}(p_1,p_2;q)
&=& {\bf q\cdot \hat{n}_{\it S}}{\Omega \over V}\sum_R V_{S;R}(q)~
\Gamma _{R;T}(p_1,p_2;q) \nonumber \\
&+& \delta_{ S,T}\Big{[} G_F^{-1}( T;p_2)-G_F^{-1}( T;p_1)
\Big{]}
\label{I1}
\end{eqnarray}
and
\begin{eqnarray}
(\nu -{\bf q\cdot v^*}_S)\Pi _{S;Q}(q)
&=&
{\bf q\cdot \hat{n}}_S{\Omega \over V}\sum_T \Pi _{T;Q}(q)~V_{Q;T}(q)+
{\bf q\cdot \hat{n}}_S{\Omega \over V}\delta_{ S,Q}
\label{I2}
\end{eqnarray}
where
\begin{eqnarray}
V_{S;T}(q)=\Bigg{(}{\bf q\times v_{\it S}^* \over |q|}\Bigg{)}~
{1\over K^0_{TT}(q)- \rho_f / m^*}
{}~\Bigg{(}{\bf q\times v_{\it T}^*\over |q|}\Bigg{)}~
\end{eqnarray}
and $K^0_{TT}(q)$ is the kernel of the bare gauge action.
In addition, we make use of the Dyson equation which relates
the reducible current vertices, $\Gamma_{S;T}$, and
the irreducible current vertices, $\Lambda_{S;T}$
\begin{eqnarray}
\Gamma _{S;T}(p_1,p_2;q) &=&
\Lambda_{S;T}(p_1,p_2;q)+\sum_{R,Q} \Pi _{S;T}(q)~\Lambda_{R;T}(p_1,p_2;q)
{}~V_{Q;R}(q)~ .
\label{I3}
\end{eqnarray}
In all the above equations, $q$ which denotes $(\nu ,{\bf q})$
is an index appearing in the bosonic quantities and $p_i$ denotes
$(\omega_i, {\bf p}_i)$, an index for the fermion variables and for example,
$G_F(S; p) \equiv G_F(\omega, {\bf k_{\it S} + p})$ with
$|{\bf k_{\it S}}| = k_F \gg |{\bf p}|$.
Combining Eqs. (\ref{I1}, \ref{I2}, \ref{I3}) gives the familiar Ward Identity
relating the current and density vertices
\begin{equation}
\nu~\Lambda_S^0(p;q)-{\bf q\cdot  \Lambda}_S(p;q)
=G_F^{-1}(S;p)-G_F^{-1}(S;p-q)
\label{general}
\end{equation}
where
\begin{equation}
\Lambda_S^0(p; q) \equiv \sum_T \Lambda_{T;S}(p - q, p; q)
\end{equation}
and
\begin{equation}
{\bf \Lambda}_S(p; q) \equiv \sum_T {\bf v}^*_T \Lambda_{T;S}(p - q, p; q)
{}~.
\end{equation}
In addition, Eqs. (\ref{I1}, \ref{I2}, \ref{I3}) can be solved directly
to determine $\Pi_{S;Q}(q)$:
\begin{eqnarray}
\Pi_{S;Q}(q) &=&{\Omega \over V}~{{\bf q\cdot \hat{n}_{\it S}}\over
\nu -{\bf q\cdot v}^*_S}\Big{[} \delta_{S,Q}+
{\Omega \over V}~{{\bf q\cdot \hat{n}_{\it Q}}\over
\nu -{\bf q\cdot v}^*_Q}~{\bf q\times v^*_{\it S}\over |q|}~
{1\over K^0_{TT}(q)- \rho_f / m^*
-v^{*2}_F\chi_t(q)}~{\bf q\times v^*_{\it Q}\over |q|} \Big{]}
\label{ASQ}
\end{eqnarray}
The diagonal elements of the correlation function are simply related to
the boson Green's function Eq. (\ref{Gaa}). These equations also
provide an explicit calculation of the irreducible vertex of the
bosonized interacting fermion system
\begin{eqnarray}
\Lambda_{S;T}(p_1,p_2;q) &=&{\delta_{S,T} \over \nu -
{\bf q\cdot v}^*_S}\Big{[} G_F^{-1}(S;p_2)-G_F^{-1}(S;p_1)
\Big{]}
\label{vertex}
\end{eqnarray}
from which it follows immediately that
\begin{eqnarray}
{\bf \Lambda}_T(p_2;q) &\equiv & \sum_S {\bf v_{\it S}^*}~
\Lambda_{S;T}(p_1,p_2;q)
= {\bf v_{\it T}^*}\Lambda_T^0
(p_2;q)
\label{identity}
\end{eqnarray}
an exact result within bosonization but in general only approximately true.
Introducing the quantity
\begin{eqnarray}
Y_S(p;q)&\equiv&
{{\bf q\cdot[\Lambda}_S(p;q)-{\bf v_{\it S}^*}~\Lambda_S^0(p;q)]\over
\Lambda_S^0(p;q) }
\label{Y}
\end{eqnarray}
and making use of the Dyson equation for the fermion self-energy
and Eq. (\ref{general}), Castellani, {\it et al.}\cite{Castellani}
found the fermion Green's function to be given by the
solution of
\begin{eqnarray}
G_F(S;p) &=& G_{F0}(S;p) + iG_{F0}(S;p)
\int {d^2q~d\nu \over (2\pi )^3}~{G_F(S;p-q)\over
\nu - v_F^*~ q_{\parallel}-Y_S(p;q)}~
D_{\mu \nu}(q)~\epsilon ^{\mu}(S;q)~\epsilon ^{\nu}(S;-q)
\label{green}
\end{eqnarray}
where
$G_{F0}^{-1}(S;p)\equiv \omega -{\bf p\cdot v}^*_S$.
When $Y_S(p;q)=0$, Eq. (\ref{green}) can be solved and the result is
exactly that derived by bosonization.

Direct calculation of $Y_S(p;q)$ allows the investigation of the
quality of approximation implicit in the bosonization approach.
We evaluate $Y_S(p;q)$ in the fermion basis, to the lowest order,
for a fermion liquid interacting via the Chern-Simons gauge field
and a two-body longitudinal interaction of the form
$V({\bf q})= g/|{\bf q}|^{1-y}$ and hence
\begin{equation}
D_{\mu \nu}(q)~\epsilon ^{\mu}(S;q)~\epsilon ^{\nu}(S;-q) \approx
v_F^{*2} {|{\bf q\times \hat{n}}_S|^2\over |{\bf q}|^2}~
{1\over i\gamma {|\nu |\over |{\bf q}|}-\chi |{\bf q}|^{1+y}}
\end{equation}
where $\gamma = {\rho _f / \pi v_F^* N^*(0)}$
and $\chi = {g / (2\pi \tilde{\phi})^2}$.
When the quadratic terms in the free particle dispersion are retained, the
bare single-particle Green's function is given by
$G_{F0}^{-1}(S;p)\equiv \omega -{\bf p\cdot v}^*_S-{p_{\perp}^2 / 2m^*}$.
Correspondingly, the bare vertex function is given by
\begin{eqnarray}
\Lambda ^0_S(p;q)&=& 1 \nonumber \\
{\bf \Lambda}_S(p;q) &=& {\bf v_{\it S}^*}
- {{\bf q}\over 2m^*}+{{\bf p}\over m^*}
\end{eqnarray}
and therefore $Y_S(p;q)$ is nonzero at leading order
\begin{eqnarray}
Y_S^0(p;q) &=& -{q_{\perp}^2\over 2m^*}+{p_{\perp}q_{\perp}\over m^*} ~ .
\label{Ycurv}
\end{eqnarray}
once curvature is accounted for.
However, it is easily seen that these terms have no effect  on the
Green's function determined from Eq. (\ref{green}). Integrating
over $q_{\parallel}$, $Y_S^0(p;q)$ exactly cancels the contributions
due to curvature in the Green's functions of the internal loop
and the solution is as before.
It remains to be shown whether or not vertex corrections due to
the transverse gauge interaction modify this result.

We follow Kim {\it et al.}\cite{Furusaki} and use a
perturbative $1/N$ expansion, where $N$ is the number of fermion flavors.
The diagrams giving the correction to the vertex function
at leading order in $1/N$ are shown in Figs.\
\ref{vertex1}, \ref{vertex2}.
At this order because the singular contributions arising from the two
diagrams of Fig.\ \ref{vertex2} cancel each other by Furry's theorem,
the only singular correction to $Y_S(p;q)$ is due
to the diagram of Fig.\ \ref{vertex1} as can be verified by direct calculation.
Furry's theorem states that for
systems with charge-conjugation symmetry, all diagrams
containing a fermion loop with an odd number of vertices may be
omitted\cite{Furry}.  Charge-conjugation (C) symmetry here is equivalent to
particle-hole symmetry.  The nonlinear pieces of the fermion dispersion,
due to curvature of the Fermi surface, break this symmetry and the
two diagrams of Fig.\ \ref{vertex2} do not cancel completely.
The remaining terms, however, are subleading and non-singular.  From
Eq. (\ref{green},\ref{Ycurv}), it is clear that the most important effect
of finite $Y_S(p;q)$ on $G_F$ comes from the pole at
$\nu-v_F^*~ q_{\parallel}+{q_{\perp}^2 / 2m^*}
-{p_{\perp}q_{\perp} / m^*} = 0$.
At the pole, the density vertex is uncorrected and the only correction
is to the transverse component of the current vertex:
\begin{eqnarray}
{\bf \Lambda}_S(p;q)|_{\rm pole}
&\approx & {\bf v}_S^*- {{\bf q}\over 2m^*}+{{\bf p}\over m^*}
+{\bf \hat{n}_{\perp}} \delta \Lambda _{S\perp }(p;q)|_{\rm pole}
\end{eqnarray}
where
\begin{eqnarray}
\delta \Lambda _{S\perp }({\bf p}, i\omega;{\bf q} ,i\nu )|_{\rm pole}
&\approx &
i{1\over q_{\perp}} {m^* v_F^* \over \gamma \pi ^2}
\int ^{\lambda /2}_{-\lambda /2} dq^{\prime}~|q^{\prime}|
\bigg{[} {\rm sgn}(\omega -\nu ) \ln \Big{(}1+ {|\omega -\nu |\gamma \over
|q^{\prime}|^{2+y}\chi } \Big{)} \nonumber \\
&&-{\rm sgn}(\omega ) \ln \Big{(}1+ {|\omega |\gamma
\over |q^{\prime}|^{2+y}\chi } \Big{)}
\bigg{]} ~ .
\label{Lperp}
\end{eqnarray}
We see from Eq. (\ref{Lperp}) that $q_{\perp}\delta \Lambda _{S\perp }$
scales as $(|\omega| \gamma / \chi)^{2/(2+y)}$;
substituted in Eq. (\ref{Y}) this results in subleading corrections
to the bare external fermion frequency, $\omega$, if $y<0$ and the fermion
Green's function is unchanged. On the other hand, if $y>0$,
the corrections dominate and the results derived from bosonization
are not reliable, as we had concluded for different reasons in
Sec. \ref{sec:transverse}.

We now consider the marginal and physically interesting case of
the longitudinal Coulomb interaction $V({\bf q})= {2\pi e^2 / |{\bf q}|}$,
$y = 0$, in detail.  Carrying out the integration in Eq. (\ref{Lperp}),
\begin{eqnarray}
\delta \Lambda_{S\perp}(p;q)|_{\rm pole} &=&
{1\over q_{\perp}}(4\zeta )\Big{[} \omega \ln \Big{(}
-i{|\omega |\over v_F^* \delta }\Big{)}
-(\omega -\nu )\ln \Big{(}-i{|\omega -\nu |\over v_F^* \delta }\Big{)}
\Big{]}+ {\rm subleading~terms} ~ ,
\end{eqnarray}
substituting in Eq. (\ref{green}) and integrating over $q_{\parallel}$
we find
\begin{eqnarray}
G_F(S;p) &=& G_{F 0}(S;p)~ \bigg{\{}
1+ {1\over (2\pi)^2}\int ^{|\omega |}_{-|\omega |}d\nu
\int dq_{\perp}~\theta (\nu \omega)~{\rm sgn}(\nu )
{{v_F^* }\over{i \gamma {{ |\omega|}\over{ |q_{\perp}|}}
- \chi |q_{\perp}| }}
\nonumber \\
&\times& {1\over G_{F0}^{-1}(S;p)
+4\zeta \Big{[} \omega \ln \big{(} -i{|\omega |\over v_F^*\delta }
\big{)}
-(\omega -\nu )\ln \big{(} -i{|\omega -\nu |\over v_F^* \delta }\big{)}
\Big{]}
-\Sigma(\omega -\nu) }
\bigg{\}}
\end{eqnarray}
The $q_{\perp}$ integral over the gauge propagator is logarithmically
singular at $\nu =0$ and the $\nu $-dependence in the denominator
may be  safely neglected on evaluating the final frequency integral.
We then recover
\begin{equation}
G_F(S;p) = G_{F0}(S;p)~\Big{[}1+\Sigma(S;p)~G_F(S;p)\Big{]}
\end{equation}
where
\begin{equation}
\Sigma (\omega )=-\zeta ~\omega ~\ln \Big{(}
{iv_F^* \delta +|\omega |\over |\omega |} \Big{)} ~ ,
\end{equation}
the marginal Fermi liquid form in Eq. (\ref{mflsigma}).

In summary, for a fermion liquid interacting via the Chern-Simons gauge
interaction and a long-range Coulomb interaction, the form of the
propagator is not modified by quadratic terms in the quasiparticle dispersion
due to curvature of the Fermi surface or by
vertex corrections beyond those already incorporated in multidimensional
bosonization.  For this system bosonization provides a consistent calculational
framework.

\section{Density Response Functions}
\label{sec:dens}

In previous sections we focused attention exclusively on the
single-particle properties of the interacting fermion system.
In particular we found evidence of non-Fermi liquid behavior in the
single-particle Green's function for the case of transverse gauge interactions.
However, the single-particle Green's function does not have clear physical
meaning as it is not gauge invariant.  Indeed, as shown by
Kim {\it et al.} \cite{Furusaki}, different types of singularities might
appear depending on the choice of gauge.  These authors used a perturbative
expansion
to study such gauge invariant quantities as the density and current response
at small momentum transfer. They were able to show that these properties
of the interacting gauge system were indeed
Fermi liquid like: singular corrections to the quasiparticle self-energy
were canceled by singular vertex corrections.
However, the possibility that singular behavior
could arise at large momentum transfer was not ruled out.  In fact, other
workers have found singular behavior at large momentum
transfer\cite{Khvesh,IKL}.  In particular, Ioffe {\it et al.}
\cite{IKL} have argued that the vertex correction is
singular at momentum transfer $2k_F$ because the gauge fields
mediate an attractive interaction between currents moving in the same
direction.  In this section, we calculate the gauge invariant response
functions
for both small momentum transfer $Q \approx 0$
and for $Q \approx 2k_F$.  First we check
that bosonization yields correct results for the free problem in both cases,
and then we take into account the effects of interactions.

We begin by writing the fermion density-density correlation
function in bosonized form.  The fermion density fluctuation is given by:
\begin{eqnarray}
\delta \rho({\bf x},t) &=&
: \psi^{\dag}({\bf x},t)~\psi({\bf x},t):
\nonumber \\
&=& \sum_{\bf S,T} : \psi^{\dag}({\bf S;x},t)~\psi({\bf T;x},t):
\end{eqnarray}
where the colons denote normal ordering which in this case is equivalent to the
subtraction of the uniform fermion background density.
Then the density-density correlation function can be written as:
\begin{eqnarray}
\langle \delta \rho({\bf x},t)~\delta \rho({\bf 0},0) \rangle
&=&
{1\over V^2}\sum_{\bf S,T} \langle J({\bf S;x},t)~J({\bf T;0},0) \rangle
+ \big{(} {\Omega \over V a}\big{)}^2 \sum_{\bf S \not = T,
U \not = V} e^{i{\bf (k_S-k_T)\cdot x}}~
\nonumber \\
&\times& \bigg{\langle} \exp \Big{\{} i {\sqrt{4\pi }\over \Omega}\big{[}
\phi({\bf S;x},t)-\phi({\bf T;x},t) \big{]}\Big{\}}~
\exp \Big{\{} i {\sqrt{4\pi }\over \Omega}\big{[}
\phi({\bf U;0},0)-\phi({\bf V;0},0) \big{]}\Big{\}}
\bigg{\rangle}
\label{fullDD}
\end{eqnarray}
For small momenta transfer, $|\bf Q| < \lambda \ll k_F$, and only
the first term in Eq. (\ref{fullDD}) is relevant.
The second term is important for momentum transfers near $Q=2k_F$.
In subsections \ref{subsec:smallq} and \ref{subsec:largeq}, we
examine the first and the second terms respectively.  We concentrate
primarily on the case of two spatial dimensions.

\subsection{Density Response at Small-$Q$}
\label{subsec:smallq}

\subsubsection{Free Fermions and Landau Fermi Liquids}

For free fermions, currents in different patches are uncorrelated.
In two dimensions, and for $|{\bf Q}| \ll k_F$,
the density response is given by the retarded correlation function
\begin{eqnarray}
\Pi^0_{00}({\bf Q},\omega )
&=& -i\int d^2x \int_0^{\infty} dt~e^{i({\bf q\cdot x}-\omega t)}
{1\over V^2}\sum_{S}\langle J(S; {\bf x}, t)~
J(S; {\bf 0}, 0) \rangle \nonumber \\
&=& N^*(0)~ \Big{(} {x\over \sqrt{x^2-1}}-1 \Big{)},
\label{pi00}
\end{eqnarray}
where $x \equiv {\omega / (v_F^* |{\bf Q}|)}$.
This is the well-known result for the
density response of the free electron gas
in the limit of low frequency and small momentum.

Including a general longitudinal
interaction $V({\bf Q})$, we use the generating functional to calculate
the correlation
between currents in different patches. Then the retarded correlation function
which combines contributions from all possible two-point boson Green's
functions
$\langle a( S;{\bf Q},\omega)~a^*( T;{\bf Q},\omega)\rangle$,
$\langle a( S;{\bf Q},\omega)~a( T;-{\bf Q},-\omega)\rangle$,
and
$\langle a^*(S;{\bf Q},\omega)~a^*( T;{\bf Q},\omega)\rangle$
is given by
\begin{eqnarray}
\langle J(S;{\bf Q},\omega)~J(T;{\bf -Q},-\omega)\rangle _{\rm Ret}
&=&
i\Omega {{\bf \hat{n}_{\it S}\cdot Q}~\delta_{S,T}\over \omega - v_F^*
{\bf \hat{n}_{\it S}\cdot Q}+i\eta } \nonumber \\
&+i& \Omega {\Lambda \over (2\pi)^2}~
{{({\bf \hat{n}_{\it S}\cdot Q})~ ({\bf \hat{n}_{\it T}\cdot Q})}\over
{(\omega - v_F^* {\bf \hat{n}_{\it S}\cdot Q}+i\eta)
(\omega -v_F^* {\bf \hat{n}_{\it T}\cdot Q}+i\eta)}}~ \nonumber \\
&& \times
{V({\bf Q})\over 1-V({\bf Q})~\Pi ^0_{00}({\bf Q},\omega)}\ .
\end{eqnarray}
Summing over $S$ and $T$, we obtain
\begin{equation}
\Pi_{00}({\bf Q},\omega) = {\Pi ^0_{00}({\bf Q},\omega)\over
1-V({\bf Q})~\Pi ^0_{00}({\bf Q},\omega)} ~ ,
\end{equation}
identical to
the RPA result for the density response which is exact in the limit
$Q, \omega \rightarrow 0$.

\subsubsection{Gauge Theory}

At two-loop order in a $1/N$ pertubative expansion,
Kim {\it et al.}\cite{Furusaki} found that beyond RPA there were only
subleading corrections to the irreducible polarizability.
We showed in Sec. \ref{subsec:ward} that implicit in the bosonization
scheme is a relation between the density and current vertices,
${\bf \Lambda}_S = {\bf v_{\it S}^*} \Lambda_S ^0$.  This relation
suggests\cite{Castellani} that within bosonization there are no
corrections to the polarizability $\Pi$ beyond those of RPA.
This supposition
is correct as we will now demonstrate by explicit calculation for the case
of the Chern-Simons gauge theory with a longitudinal interaction.

The correlation function between currents in different patches can be
found using the generating functional and is given by
\begin{eqnarray}
\langle J(S;{\bf Q},\omega)~J(T;{\bf -Q},-\omega)\rangle _{\rm Ret}
&=& i\Omega {{\bf \hat{n}_{\it S}\cdot Q}~\delta_{S,T}\over \omega -v_F^*
{\bf \hat{n}_{\it S}\cdot Q}+i\eta} \nonumber
\\&+& i \Omega {\Lambda \over (2\pi)^2}~
{{({\bf \hat{n}_{\it S}\cdot Q})~ ({\bf \hat{n}_{\it T}\cdot Q})}\over
{(\omega -v_F^* {\bf \hat{n}_{\it S}\cdot Q}+ i\eta)
(\omega -v_F^* {\bf \hat{n}_{\it T}\cdot Q}+i\eta)}}
\nonumber \\
&\times& \bigg{[}D^R_{00}({\bf Q},\omega)+
{\bf Q\times v_{\it S}^*\over |Q|}~D^R_{T0}({\bf Q},\omega)
-{\bf Q\times v_{\it T}^*\over |Q|}~D^R_{0T}({\bf Q},\omega) \nonumber
\\ &-&
{\bf Q\times v_{\it S}^*\over |Q|}~{\bf Q\times v_{\it T}^*\over |Q|}~
D^R_{TT}({\bf Q},\omega) \bigg{]}
\end{eqnarray}
where $D^R_{\mu \nu}$ denotes the retarded part of the gauge propagator.
Only the term proportional to $D^R_{00}({\bf Q},\omega)$
survives the sum over patches
and we obtain for the density response:
\begin{eqnarray}
\Pi_{00}({\bf Q},\omega) &=&
\Pi^0_{00}({\bf Q},\omega)\big{[}1+D^R_{00}({\bf Q},\omega)
{}~\Pi^0_{00}({\bf Q},\omega)\big{]}
\nonumber \\
&=& {{Q^2/ (2\pi \tilde{\phi})^2}\over{{Q^2 / (2\pi \tilde{\phi})^2}
-\Pi^0_{00}({\bf Q},\omega)\big{[}{\rho_f / m^*} + Q^2 V({\bf Q})
- \Pi^0_{11}({\bf Q},\omega)\big{]}}}
\end{eqnarray}
where $\Pi^0_{11}({\bf Q},\omega)$ is the current response function
of the free electron gas
\begin{equation}
\Pi^0_{11}({\bf Q},\omega)
=-v_F^{*2} N^*(0)~ \big{[}x^2-{1\over 2}-x\sqrt{x^2-1} \big{]} ~ .
\end{equation}
In the same way, we may evaluate the current response
function:
\begin{eqnarray}
\Pi_{11}({\bf Q},\omega) &=& \Pi^0_{11}({\bf Q},\omega)
\big{[}1-D^R_{TT}({\bf Q},\omega)~\Pi^0_{11}({\bf Q},\omega)\big{]}
\end{eqnarray}
where
\begin{eqnarray}
D^R_{TT}({\bf Q},\omega) &=&
{\Pi^0_{00}({\bf Q},\omega)\over{\Pi^0_{00}({\bf Q},\omega)
\big{[} \rho_f / m^* + Q^2 V({\bf Q}) -
\Pi^0_{11}({\bf Q},\omega)\big{]} - Q^2 / (2\pi \tilde{\phi})^2}}
\end{eqnarray}
As stated above, we reproduce the RPA results, in agreement with the
conclusion of Kim {\it et al.}

\subsection{Density Response at $Q \approx 2 k_F$}
\label{subsec:largeq}

\subsubsection{Free Fermions and Landau Fermi Liquids}

It is important to check whether bosonization captures the physics of
the Kohn anomaly\cite{Kohna}.  At first glance it is not
obvious how bosonization, with the restriction $|{\bf q}| < \lambda \ll k_F$,
can possibly describe processes with large momentum transfer, especially those
near $2 k_F$.  The reason why it can
is made clear by observing that scattering between {\it different}
patches is allowed, and that in fact the second term in
Eq. (\ref{fullDD}) contains these processes explicitly.
Recall that $2 k_F$ processes are described correctly by standard
one-dimensional bosonization\cite{Peschel,Luther}.  Bosonization in one and
higher spatial dimensions correctly describes processes with either small
or large momentum, provided they are of low energy.  That is why the
specific heat, computed within the bosonization framework,
is exact in the low-temperature limit\cite{Haldane,Tony}.

Since for free fermions the bosons in different patches are independent,
the correlation function in the second term of Eq. (\ref{fullDD})
is zero except when $\bf V = S$ and $\bf U = T$, and the $Q \approx 2k_F$
contribution to the density response function is given by:
\begin{eqnarray}
\Pi ^0_{2k_F}({\bf x},t) &=& -i \big{(} {\Omega \over V a}\big{)}^2
\sum_{\bf S \not = T} e^{i{\bf (k_S-k_T)\cdot x}}~
\bigg{\langle} e^{i {\sqrt{4\pi }\over \Omega}\big{[}
\phi({\bf S;x},t)-\phi({\bf T;x},t) \big{]}}~
e^{i {\sqrt{4\pi }\over \Omega}\big{[}
\phi({\bf T;0},0)-\phi({\bf S;0},0) \big{]}} \bigg{\rangle} ~\theta(t)
\nonumber \\
&=& i\bigg{[} {{\Lambda^{D-1}}\over{(2 \pi)^D}}~ \sum_{\bf S}
{{e^{i{\bf k_S \cdot x}}}\over{{\bf \hat{n}_S \cdot x} -v^*_F t +
i \eta~ {\rm sgn}(t)}}
\bigg{]} \bigg{[} {{\Lambda^{D-1}}\over{(2 \pi)^D}}~ \sum_{\bf T}
{{e^{-i{\bf k_T \cdot x}}}\over{{\bf \hat{n}_T \cdot x} - v^*_Ft +
i \eta~ {\rm sgn}(t)}} \bigg{]}~\theta(t) ~ .
\label{largeq}
\end{eqnarray}
Implicit in the last line is the restriction $|{\bf x}_\perp \Lambda| < 1$.
As we are interested in the limit $Q \rightarrow
2k_F$, only patches which are nearly opposite to each other on the Fermi
surface
participate in the scattering, ${\bf \hat{n}_S \approx - \hat{n}_T}$.
Therefore we may carry out the sum over patches in each of the two bracketed
terms in Eq. (\ref{largeq}), keeping in mind that the spatial directions
perpendicular to the Fermi surface normal are the same for both terms.
To simplify the calculation, we consider only the static ($\omega = 0$)
response and specialize to the case of a circular Fermi surface in
two dimensions.  Integrating the  density response,
Eq. (\ref{largeq}), over positive time we obtain:
\begin{equation}
\Pi ^0_{2k_F}(R, \omega =0) = -{k_F^2\pi \over v^*_F(2\pi )^4}
\int_0^{2 \pi} d \theta_1 d \theta_2~ {{\exp[i k_F R~ (\cos\theta_1
- \cos\theta_2)] [{\rm sgn}(\cos\theta_1) - {\rm sgn}(\cos\theta_2)]}\over{
(\cos\theta_1 - \cos\theta_2)~ R}}
\label{dens2}
\end{equation}
where $\theta_{1,2}$ are the angles between $\bf x$ and
${\bf k}_S$ and ${\bf k}_T$ respectively
and $R = |{\bf x}|$.  Now we may focus on the
case of $Q \approx 2k_F$ by limiting the range of integration over the
angular variables $\theta_{1,2}$ to angles near $\theta_1 = 0$
and $\theta_2 = \pi$.  In this limited region the cosines appearing
in Eq. (\ref{dens2}) may be expanded in Taylor-series as $\cos(\theta_{1,2}) =
\pm (1 - \theta_{1,2}^2/2)$ and the Gaussian integrals over the angles can be
carried out.
Finally, the Fourier transform into $Q$-space is carried out.  If $\theta$ is
the angle between $\bf Q$ and $\bf x$, the density response function is given
by:
\begin{eqnarray}
\Pi ^0_{2k_F}(Q, \omega =0)
&=& {2i~k_F\pi ^2\over v^*_F(2\pi)^4} \int_0^\infty R dR \int_0^{2 \pi}
d\theta {{\exp[i (2 k_F - Q \cos\theta) R~]}\over{R^2}}
\nonumber \\
&=& {1\over 8}~{k_F^{1/2}\over v^*_F\pi ^{3/2}}e^{i\pi /4}
 \int_a^\infty dR~{{\exp[i (2 k_F - Q) R~]}\over{R^{3/2}}}~ .
\label{dens3}
\end{eqnarray}
In the second line we have assumed $Q \geq 2 k_F$
and have also introduced an ultraviolet
cutoff $a$ to eliminate the artificial
divergence at small-$R$.  The nonanalytic
term we seek comes from the large-$R$ region defined by $k_F R \gg 1$.
Again in this limit a Taylor-series expansion of the $\cos \theta $
term is justified and the resulting Gaussian integral in the $\theta$
variable has been carried out.  For $Q < 2 k_F$, obviously
$Q \cos\theta < 2k_F$ and consequently the integral over $\theta$ contains only
terms which oscillate rapidly with increasing $R$ and which therefore do not
generate non-analytic behavior at large-$R$.  Finally, asymptotic analysis of
the radial integral gives the non-analytic term in the density response
function:
\begin{eqnarray}
\Pi ^0_{2k_F}(Q, \omega=0) &=& {{\sqrt{2} m^*}\over 4\pi }~
\sqrt{{{Q-2k_F}\over{2k_F}}}+ C;\ Q > 2k_F
\nonumber \\
&=& C;\ Q < 2k_F\ .
\label{2d2kf}
\end{eqnarray}
Here $C$ is a constant which can be determined by a more careful evaluation
of the integrals in the small-$R$ region.  Continuity across $Q = 2k_F$
ensures that the same constant appears in both limits of Eq. (\ref{2d2kf}).
The Kohn anomaly of Eq. (\ref{2d2kf}) agrees precisely with that found
directly in the fermion basis.\cite{Stern}
Repeating the calculation in $D=3$ we easily show that the static
response is given by:
\begin{equation}
\Pi ^0_{2k_F}(Q, \omega =0)
={{k_Fm^*}\over{2 (2 \pi)^2}}~ {{2k_F - Q}\over Q}\big{[} \ln {{|Q-2 k_F|}
\over{2 k_F}} + \gamma -1-{i\pi \over 2}\big{]} + C^\prime
\label{3d2kf}
\end{equation}
which again agrees with the exact result found in the fermion basis\cite{Kohna}
in the limit $Q \rightarrow 2 k_F$.
The Kohn anomaly persists even if longitudinal interactions are included.
In fact, this has already been anticipated in Eq. (\ref{2d2kf}) and
Eq. (\ref{3d2kf}) where the effective mass
$m^*$ rather than the bare mass $m$ appears.

\subsubsection{Gauge Theory}

Now we consider the real-space density response function in two dimension
at momentum transfer $2k_F$ for the Chern-Simons
theory.  With the gauge interaction, currents moving in the same direction are
attracted to each other so we might anticipate a singularity in
$\Pi _{2k_F}$.  This singularity has a different physical origin than the
Kohn anomaly of the previous section, which was not singular, just nonanalytic.
The Kohn anomaly is due to the reduced availability of the phase space
at low energies.
In contrast, the singularity, if it exists, should be apparent even
if we consider only scattering between two patches at exact opposite points of
the Fermi surface.
Therefore we need consider only terms with ${\bf k_{\it S}=-k_{\it T}}$
in the $2k_F$ response function, Eq. (\ref{largeq}).
\begin{eqnarray}
\Pi_{2k_F}({\bf x},t) &=&
-i \big{(} {\Omega \over V~a} \big{)}^2 \sum_{ S} \exp \Big{\{}
{4\pi \over \Omega ^2} \big{[} \langle \phi({\bf {\it S}; x},t)
\phi({\bf {\it S}; 0},0)
\rangle +\langle \phi({\bf {\it -S}; x},t)\phi({\bf {\it -S}; 0},0)\rangle
\nonumber \\
&-& \langle \phi({\bf {\it S}; x},t)\phi({\bf {\it -S}; 0},0)\rangle
-\langle \phi({\bf {\it -S}; x},t)\phi({\bf {\it S}; 0},0)\rangle \big{]}
\nonumber \\
&-&{4\pi \over \Omega ^2} \big{[} {\rm terms~ with~ }
({\bf x},t) \rightarrow ({\bf 0},0) \big{]} \Big{\}}~\theta(t) ~ .
\label{2kf}
\end{eqnarray}
The in-patch boson correlation function is given by Eq. (\ref{Gaa}):
\begin{eqnarray}
\langle \phi({\bf {\it S}; x},t)\phi({\bf {\it S}; 0},0)
-\phi^2({\bf {\it S}; 0},0)\rangle
&=& {\Omega ^2 \over 4\pi} \ln \bigg{(} {ia\over {\bf x\cdot \hat{n}}_S
- v_F^* t}
\bigg{)} \nonumber \\
&+& {\Omega ^2 \over 4\pi}i \int {d^2 q\over (2\pi)^2}\int {d\omega \over 2\pi}
[ e^{i({\bf q\cdot x}-\omega t)}-1]{1\over (\omega -
v_F^*~ {\bf q\cdot \hat{n}}_S)^2} \nonumber \\
&\times& \bigg{\{} D^R_{00}(q)-v_F^{2*}
{|{\bf q\times \hat{n}}_S|^2\over |{\bf q}|^2} D^R_{TT}(q) \bigg{\}}~ .
\label{2kf1}
\end{eqnarray}
The correlation function between bosons at opposite points of the
Fermi surface contains the new physics.  The correlation function
$\langle \phi({\bf {\it S}; x},t)\phi({\bf {\it -S}; 0},0)
-\phi({\bf {\it S}; 0},0)
\phi({\bf {\it -S}; 0},0)\rangle$
can be found using the generating functional and is given by an integral
over frequency and momenta of:
\begin{eqnarray}
&&{\theta(q_\parallel) \over {q_\parallel}} \bigg{\{} \big{(}
e^{i(q_{\parallel}
x_{\parallel}-\omega t)}-1 \big{)}\langle a({\bf {\it S};q})~
a({\bf {\it -S}; -q})\rangle
+\big{(} e^{-i(q_{\parallel} x_{\parallel}-\omega t)}-1 \big{)} \langle
a^{\dag}({\bf {\it S}; q})~ a^{\dag}({\bf {\it -S}; -q})\rangle \bigg{\}}
\nonumber \\
&=& i {\Omega \over V}
{{\theta(q_\parallel)}\over{v_F^{2*} q_{\parallel}^2-\omega ^2}}
\bigg{\{} \big{(} e^{i(q_{\parallel} x_{\parallel}
-\omega t)}-1 \big{)}\Big{[} D^R_{00}(q)+v_F^* {q_\perp \over |q|}
\{ D^R_{0T}(q)+D^R_{T0}(q) \}+{v_F^{2*}
q_{\perp}^2\over |{\bf q}|^2} D^R_{TT}(q)\Big{]}
\nonumber \\
&&+\big{(} e^{-i(q_{\parallel} x_{\parallel}-\omega t)}-1 \big{)}\Big{[}
D^R_{00}(q)-v_F^* {q_\perp
\over |q|}\{ D^R_{0T}(q)+D^R_{T0}(q) \}
+{v_F^{2*} q_{\perp}^2\over |{\bf q}|^2} D^R_{TT}(q)\Big{]}\bigg{\}}~ .
\label{2kf2}
\end{eqnarray}
As in the case of the two-point function,
the most singular contribution comes from the $q$-limit.
Inserting Eqs. (\ref{2kf1}, \ref{2kf2}) into Eq. (\ref{2kf}) and computing
the integrals in $q$-limit we obtain
\begin{eqnarray}
\Pi_{2k_F}({\bf x},t) &=&  {\sum_{S}}^\prime
{-i\over \big{[}{\bf x\cdot \hat{n}}_S-v_F^* t
+i\eta ~{\rm sgn}({\bf x\cdot \hat{n}}_S)\big{]}~
\big{[}{\bf x\cdot \hat{n}}_S+v_F^* t
+i\eta ~{\rm sgn}({\bf x\cdot \hat{n}}_S)\big{]}}
{}~\exp \big{[} {\cal{E}}({\bf x\cdot \hat{n}}_S,t) \big{]}
\nonumber \\
&\propto& {-i\over \big{(}|{\bf x}|-v_F^* t + i\eta \big{)} \big{(}|{\bf x}|
+ v_F^* t + i\eta \big{)}}
{}~\exp \big{[} {\cal{E}}(|{\bf x}|,t) \big{]} ~ ,
\end{eqnarray}
where the prime on the above sum over patches indicates that the sum
is only over patches for which
$|{\bf x \times \hat{n}}_S| \Lambda < 1$ and where
\begin{eqnarray}
{\cal{E}}(x,t) &\approx &
-\zeta \Big{\{} {x\over x-v_F^* t +i\eta
{}~{\rm sgn}(x)}\ln \big{|} (x-v_F^* t)\delta \big{|}
+{x\over x+v_F^* t +i\eta
{}~{\rm sgn}(x)}\ln \big{|} (x+v_F^* t)\delta \big{|}
\nonumber \\
&-&\ln ^2\big{|} (x-v_F^* t)\delta \big{|}
-\ln ^2\big{|} (x+v_F^* t)\delta \big{|} \Big{\}} ~ ,
\end{eqnarray}
for $|x_{\parallel}-v^*_F t| \delta \gg 1$
and $|x_{\parallel}+v^*_F t| \delta \gg 1$.
The log-squared terms appearing in ${\cal{E}}$
have their origin in the correlations induced by the gauge interaction
between the bosons in opposite hemispheres.
Evidently a log-squared singularity at momentum transfer $2k_F$ exists
in the real-space response function.
Although it is technically difficult to Fourier transform it into
frequency space, we can estimate the leading singular factor of
$\Pi_{2k_F}(\omega)$ with the use of a scaling argument and find:
\begin{equation}
\Pi_{2k_F}(\omega) \propto \exp(2 \zeta \ln ^2|\omega / (v_F^* \delta) |)
\end{equation}
where again $v_F^* \delta = O(\epsilon_F)$.
This singularity is similar to, though stronger than,
that found by Altshuler {\it et al.} \cite{Altshuler}.

\section{Transverse Gauge Fields: Fermions in Three Dimensions}
\label{sec:EM}
In this section, we discuss the most familiar physical example of fermions
interacting via gauge interactions: electrons in three spatial dimensions with
the full Maxwell electromagnetic field.
The longitudinal Coulomb interaction is screened,
and the transverse gauge interaction is suppressed by a factor of fine
structure constant multiplied by the ratio of the Fermi velocity to
the speed of light.  For these reasons we expect the behavior to be controlled
by the Landau Fermi liquid fixed point except at extremely low
temperatures.  Nevertheless it is an interesting exercise to study the
crossover to the non-Fermi liquid fixed point.

An expansion in the fine structure constant leads to a modified
quasiparticle spectrum\cite{Pethick} near the Fermi surface:
\begin{equation}
\epsilon({\bf p}) =
v_F(|{\bf p}|-p_F)-ae^2(v_F/c)^2(|{\bf p}|-p_F)\ln \big{|}|{\bf p}|-p_F\big{|}
{}~ ,
\end{equation}
here $a$ is a positive constant.
It has been conjectured that the logarithmic
term exponentiates to give a power law spectrum\cite{Pethick},
\begin{equation}
\epsilon({\bf p}) =
v_F(|{\bf p}|-p_F)\big{|}|{\bf p}|-p_F\big{|}^{-a\alpha v_F/c}~ .
\end{equation}
Here we investigate the fixed point using the formalism developed in
previous sections.  However, we do not find a
power law, rather, a logarithmic spectrum of
marginal Fermi liquid type characterizes the fixed point.

We consider the following bare gauge action in the Coulomb gauge:
\begin{equation}
S^0_G[A]={1\over 8\pi (e/c)^2}
\int d^3x~dt~\Big{[}{{1}\over{c^2}}~ ({\partial {\bf A}\over \partial t})^2
-({\bf \nabla \times A})^2\Big{]} ~ ,
\end{equation}
where the longitudinal interaction is neglected since it is
screened\cite{HKMS}. Once the bare gauge action, Eq.(\ref{SG0}), is specified
we can follow the prescription given in Sec. \ref{sec:bose} to obtain:
\begin{equation}
D_{\mu \nu}(q)~ \epsilon^{\mu}({\bf S})~ \epsilon^{\nu}({\bf S})=
{{|{\bf q\times \hat{n}_S}|^2}\over{|{\bf q}|^2}} {{4\pi v_F^{*2} e^2}\over
{4\pi e^2 ~\chi^T(q) + \omega^2 - c^2 |{\bf q}|^2}}~ ,
\label{Gmn3D}
\end{equation}
where the transverse susceptibility $\chi^T(q)$ was defined in Eq.(\ref{chiT}).
This result is in agreement with the RPA calculation of the gauge
propagator\cite{Pethick}.  In $\omega $-limit, $\omega > v_F^*|{\bf q}|$,
the poles of Eq. (\ref{Gmn3D}) are at
$\omega^2 = 4\pi \rho_f e^2 / m + c^2|{\bf q}|^2$,
the plasma frequency.
In the opposite q-limit, $\omega < v_F^* |{\bf q}|$, we obtain:
\begin{equation}
D_{\mu \nu }(q)~ \epsilon^{\mu}({\bf S})~\epsilon^{\nu}({\bf S})=
v_F^{*2} {|{\bf q\times \hat{n}_S}|^2\over |{\bf q}|^2}{1\over
i~\gamma {|\omega |\over |{\bf q}|}-\chi |{\bf q}|^2}
\label{damp3D}
\end{equation}
where now $\gamma = \pi v_F^* N^*(0)/4$ and $\chi ={c^2 / 4 \pi e^2}$.
Inserting this expression into the boson propagator Eq.(\ref{Gaa}),
we obtain the imaginary part of the boson
self-energy due to quasiparticle damping.  As in the case of the half-filled
Landau level, the $q$-limit is
the more singular of the two limits, and is the origin of the non-Fermi
liquid fixed point.  From
Eqs.(\ref{Gaa}, \ref{Gff}, \ref{damp3D}), we can now compute the real space
fermion Green's function.  Factorizing the quasiparticle Green's function
into a product of the free term
$G^0_f({\bf S;x},t)=({\bf x\cdot \hat{n}_S}-v_F^*~ t)^{-1}$
and the correction term $\exp[\delta G_B({\bf S;x},t)]$, we have
\begin{eqnarray}
\delta G_B({\bf S;x},t)&=&
i\int {d^2q_{\perp}\over (2\pi)^2}\int {dq_{\parallel}
\over 2\pi }\int {d\omega \over 2\pi }~
[e^{i({\bf q\cdot x}-\omega t)}-1]~
{D_{\mu \nu }({\bf q},\omega )~\epsilon ^\mu ({\bf S};q)
{}~\epsilon ^\nu ({\bf S};-q) \over
[\omega -v_F^*~ {\bf q\cdot \hat{n}_S}+i\eta ~{\rm sgn}(\omega )]^2 }
\nonumber \\
&\approx &i\int {d^2q_{\perp}\over (2\pi)^2}\int_0^{v_F^* \lambda / 2}
{d\omega \over 2\pi}
{}~|x_{\parallel}|~e^{i\omega ~{\rm sgn}(x_{\parallel})
{}~(x_{\parallel}/v_F^* - t)} ~{ |{\bf q_{\perp}}| \over
\chi |{\bf q_{\perp}}|^3 -i\gamma \omega }
\nonumber \\
&\approx & {4 \pi i e^2 \over 3 c^2}~ |x_{\parallel}|
\int_0^{v_F^* \lambda / 2}{d\omega \over 2\pi }
{}~e^{i\omega ~{\rm sgn}(x_{\parallel})
{}~(x_{\parallel} / v_F^*-t)}~ \ln \Big{(}
{{\lambda^3 \chi / 8 \gamma - i\omega}\over{-i\omega}} \Big{)} ~ ,
\label{int3D}
\end{eqnarray}
in the second line we use the fact that $|{\bf q_{\perp}}| \gg
|q_{\parallel}|$ in the region of integration where the integrand is the
most singular, and perform a contour integration around the pole located at
$q_{\parallel}={\omega / v_F^*}$.  The frequency
integration in the last line of Eq.(\ref{int3D})
is identical in form to Eq. (\ref{int2D}).  Introducing a new momentum scale
$\delta = \lambda^3 c^2 / 4 e^2 v_F^* k_F^2$
and dimensionless constant $\zeta = 2 e^2 v_F^* / 3 c^2$, we obtain:
\begin{eqnarray}
\delta G_B({\bf S;x},t)
&\approx& -\zeta |\delta x_{\parallel}|~\ln |
\delta (x_{\parallel}-v_F^* t) |~;~
|x_{\parallel}-v_F^* t|\delta \ll 1
\nonumber \\
&\approx &-\zeta {x_{\parallel}\over x_{\parallel}-v_F^* t + 2 i a~ {\rm sgn}
(x_{\parallel})}\Big{[} \gamma + \ln \big{|}(x_{\parallel}-v_F^* t)
\delta ~{\rm sgn}(x_{\parallel})\big{|}\Big{]};\
|x_{\parallel}-v_F^* t|\delta \gg 1
\label{3Dgreen}
\end{eqnarray}
As in the case of the half-filled Landau level, the Green's function has
the marginal Fermi liquid form with self-energy given by Eq. (\ref{mflsigma}).
A similar result was found in a fermion renormalization group calculation by
Chakravarty {\it et. al}.\cite{Norton}
However, because $\zeta \approx 10^{-5}$ for a typical metal,
the crossover to MFL behavior occurs at an exceedingly low energy scale
$\omega$ set by
\begin{equation}
\zeta~ \log(E/\omega) = 1
\end{equation}
where $E \equiv (c^2 k_F) / (4 e^2)$.
Including nonlinear terms due to curvature of the Fermi surface
in the fermion dispersion and vertex corrections due to the gauge interactions,
we may apply the arguments presented in Sec. \ref{subsec:ward}.
It is straightforward task
to show that bosonization is consistent and provides an adequate
description of the three dimensional gauge-fermion liquid.

\section{Conclusions}
\label{sec:conc}
In multidimensional bosonization, Feynman diagrams are replaced
by geometrical and algebraic ways of understanding strongly correlated
fermions. The generating functional approach is a convenient way to
compute correlation functions of bosons and hence, fermions,
interacting via short- and long-range longitudinal, and transverse gauge
interactions.  In one dimension well-known results for the Luttinger liquid
are recovered.  Short-range and Coulomb interactions in two or higher spatial
dimensions do not destroy the Landau Fermi liquid fixed point; spin-charge
separation does not occur.  The Kohn anomaly in the density response
function near momentum $2 k_F$ has the standard nonanalytic form, showing
that all low energy processes, even those with large-momentum,
are correctly described
by multidimensional bosonization.  Likewise, the specific heat has the
correct linear dependence on temperature with subleading corrections which
are easily computed.\cite{Tony}  Collective modes in the charge and spin
sectors follow directly from the equations of motion for the abelian $U(1)$
and non-abelian $SU(2)$ currents.\cite{HKM}
The stability of the Landau fixed point
against superconducting BCS, CDW, and SDW instabilities also can be ascertained
using the renormalization-group in the boson basis.\cite{HKM}
Novel fixed points in higher dimension arise in several cases.
These examples demonstrate that bosonization is more encompassing
than Landau Fermi liquid theory as non-Fermi liquid fixed points are
accessible.
For a super-long range longitudinal interaction in two dimensions,
a system first studied by Bares and Wen,\cite{Wen}
we find an anomalous exponent in the equal-time Green's function.
For transverse gauge interactions
we find marginal Fermi liquid (MFL) fixed points in two instances:
for the HLR theory of the half-filled Landau level with the Coulomb interaction
and for ordinary Maxwell gauge interactions in three dimensions.

We studied the gauge-invariant density response function at the half-filled
Landau level fixed
point.  At low momentum and frequencies we obtain the same result as the RPA
calculation, and the response function is
indistinguishable from that of a Landau Fermi liquid.  Deviations from Fermi
liquid behavior appear near momentum transfers of $2k_F$.
Since odd-denominator fractional fillings are equivalent, via Jain's
construction,\cite{Jain} to the half-filled Landau level in an
additional external magnetic field $\Delta B$,
it would be interesting to examine questions of the
divergent renormalization of the composite fermion mass\cite{Du,Leadley}
and the associated compressibility\cite{KLWS} as a function of $\Delta B$
within the bosonization framework.

The intimate connection between multidimensional bosonization and the
Ward Identity method of Castellani, Di Castro, and Metzner
enables us to go outside the framework of bosonization
to check its consistency and accuracy.  Neither non-linear terms
in the fermion dispersion due to Fermi
surface curvature, nor vertex corrections
due to gauge fields, change our results in any qualitative way.  All of the
fixed points we found possess much higher symmetry than the bare interacting
problem.  This infinite $U(1)$ symmetry reflects vanishingly small
large-angle scattering at such zero temperature quantum critical fixed points.
Multidimensional bosonization is a good description of Fermi and
nearly-Fermi liquids.

\section*{Acknowledgements}
We thank P. A. Bares, C. Castellani, C. Di Castro, J. P. Eisenstein,
B. I. Halperin, G. Kotliar, A. Ludwig,
A. Millis, W. Metzner, P. Nelson, R. Shankar, A. Stern, and C. M. Varma for
helpful discussions.  Many helpful conversations
arose at the Aspen Center for Physics where some of this work was completed.
The research was supported in part by the National Science Foundation
through grants DMR-9008239 (A.H.) and DMR-9357613 (J.B.M.) and by a grant
from the Alfred P. Sloan Foundation (J.B.M.).

\appendix
\section{Numerical Computation of the Spectral Function}

Small-angle scattering between the fermion quasiparticles generates
a subleading additive
correction to the free boson propagator.  We use the bosonization
formula Eq. (\ref{fermi}) to compute the fermion spectral function.
Since bosonization is carried out in $({\bf x}, t)$
space we must perform three operations.  First we inverse Fourier transform
the boson propagator into real space and time.
Next the exponential of the resulting expression
yields the fermion propagator.  Finally a second Fourier transform
of the fermion propagator back into momentum space allows us to extract the
spectral function.

It is a difficult technical problem to carry out these three steps
analytically.  In this Appendix we resort to the numerical method of
fast Fourier
transforms (FFTs) to compute the full spectral function.  The problem is
still challenging since we seek to extract sub-leading corrections to scaling.
For example, the imaginary part of the fermion self-energy\cite{HKM}
is of order $(\omega^2 / \epsilon_F)~ \ln(|\omega|)$ and therefore much
smaller than the
quasiparticle energy at small frequencies.  In fact, for spectral
broadening to show up in the numerical calculation, it is necessary to
set $\lambda = O(k_F)$.
As we shall see, setting $\lambda = O(k_F)$ forces us to treat
velocity renormalization carefully.

As a illustrative example of the numerical method,
we consider the problem of a two-dimensional spinless Fermi liquid with only a
single non-zero Landau parameter, $f_0$.  In this case we keep only the
temporal component of the gauge fields appearing in Eq. (\ref{Kmn})
and set $K^0_{00}(q) = 1/f_0$.  Then the gauge propagator appearing in
Eq. (\ref{fermi}) is given by:
\begin{equation}
D_{\mu \nu}({\bf q}, \omega) ~\epsilon^\mu( S; q)
{}~\epsilon^\nu( S; -q) = {f_0\over{1 + f_0~ \chi^0(q)}}\ .
\label{f0self}
\end{equation}
To continue our analysis, we now set $x_\perp = 0$ in Eq. (\ref{fermi})
as the spectral function has only weak and uninteresting
dependence on transverse momentum.  With this simplification, the integral
over $q_\perp$ in Eq. (\ref{fermi}) may be performed analytically, yielding
a rather complicated expression which, for the sake of brevity,
we do not present here\cite{code}.  Instead we denote the integral by:
\begin{equation}
D(q_\parallel, \omega) \equiv \int_{-\lambda/2}^{\lambda/2}~
{{dq_\perp}\over{2 \pi}}~ {f_0\over{1 + f_0~ \chi^0(q)}}
\label{D}
\end{equation}
which, multiplied by the double pole $[\omega - v_F^\prime~ q_\parallel +
i \eta~ {\rm sgn}(\omega)]^{-2}$,
is the quantity to be Fourier transformed into $(x_\parallel, t)$ space.
Here we distinguish between $v_F^\prime$, which is the velocity of the
non-interacting bosons, and $v_F^*$, which is the final velocity of the
bosons (and fermions) after the effect of the current-current interaction
$f_0$ on the velocity has been accounted for.

For Landau Fermi liquids, the fermion velocity renormalizes from its free
value when high-energy processes, those with energy greater than the scale
set by the cutoff $\lambda$,
are integrated out.  As long as $\lambda \ll k_F$, further renormalization of
the velocity due to the remaining low-energy processes, the ones to be
bosonized, is insignificant\cite{HKM}.
Velocity renormalization cannot, however, be neglected when
$\lambda$ is comparable to the Fermi momentum.  To incorporate this
renormalization directly into the Green's function, from
$D(q_\parallel, \omega)$ we subtract its value at $q_\parallel
= \omega = 0$, $D(0, 0)$.  Setting $F_0 \equiv f_0 N^\prime(0)$, where
$N^\prime(0) = k_F / 2 \pi v_F^\prime$, we find
$D(0, 0) = {{v_F^\prime \lambda}\over{k_F}}~ {{F_0}\over{1
+ F_0}}$.  The final renormalized Fermi velocity is then given by:
\begin{equation}
v_F^* = (1 + {{\lambda}\over{2 \pi k_F}}~ {{F_0}\over{1 + F_0}})~ v_F^\prime \
{}.
\label{shift}
\end{equation}
The difference $D(q_\parallel, \omega) - D(0, 0)$, which is small
for small frequencies and momenta, now appears in the numerator of
the remaining integral over $\omega$ and $q_\parallel$ in Eq. (\ref{fermi}):
\begin{equation}
G_F(S; x_\parallel, t) = {{e^{i k_F x_\parallel}}\over
{x_\parallel - v_F^*~ t}}~
\exp\bigg{\{} i\int {dq_\parallel\over2\pi} \int{d\omega \over 2\pi}~
{{[e^{i(q_\parallel x_\parallel - \omega t)}-1]~
[D(q_\parallel, \omega) - D(0, 0)]}\over
{[\omega - v_F^*~ q_\parallel + i\eta ~{\rm sgn}(\omega )]^2}}
\bigg{\}}\ .
\label{fermi2}
\end{equation}
These two integrals may now be performed numerically by a two-dimensional
(inverse) FFT.  In practice
it is simplest to study either the advanced or retarded Green's function rather
than the full time-ordered Green's function; hence we restrict the FFT to
the half plane $t > 0$.  Upon subtracting the
$x_\parallel = t = 0$ value, exponentiating the difference, and multiplying
by the free Green's function, the full fermion Green's function is obtained in
real space and time.  This result may then be converted back
to $(k, \omega)$
space by the second FFT.  The spectral function is then extracted by taking
the imaginary part\cite{code}.  Displayed in Fig.\ \ref{spectral1}
are cross-sectional plots of the spectral function at different frequencies
computed on a $2048 \times 2048$ lattice.  The peak has been centered in
each plot.  The width of the peak at zero frequency
is due solely to finite-size effects.  As the frequency increases from
zero, the width grows rapidly as expected.  The nonperturbative results
demonstrate that low-order perturbative expansions in the Landau parameters are
qualitatively correct as there is no breakdown in Landau Fermi liquid behavior.

\section{General Landau Parameters in Two and Three Dimensions}

It is straightforward to include higher order Landau parameters in the
computation of the 2N-point boson correlation functions.  Extension to three
spatial dimensions is straightforward.  Finally, Landau parameters in the spin
sector may be incorporated in a perturbative way
with the use of Abelian bosonization.
In two dimensions the boson propagator for a single $f_n$ interaction in the
charge sector is given by:
\begin{eqnarray}
\langle a(S;q)~a^{\dag}(S;q)\rangle &=&
{i\over \omega - v_F^* {\bf q\cdot \hat{n}}_S+i\eta ~{\rm sgn}(\omega )}
\nonumber \\
&-& i{\Lambda \over (2\pi )^2 } {\bf q\cdot \hat{n}}_S
{{\cos^2[n~\beta({\bf {\it S};q})]D^{(n)}_l(q)+
\sin^2[n~\beta({\bf {\it S};q})]
D^{(n)}_t(q)}
\over{[\omega -v_F^* {\bf q\cdot \hat{n}}_S+i\eta ~{\rm sgn}(\omega
)]^2}}~ ,
\label{fn}
\end{eqnarray}
where
\begin{eqnarray}
D^{(n)}_l(q) &=& {1\over -1/f_n + \chi ^{(n)}_l(q)} ~ ,  \\
D^{(n)}_t(q) &=& {1\over -1/f_n + \chi ^{(n)}_t(q)} ~ ,
\end{eqnarray}
and
\begin{eqnarray}
\chi ^{(n)}_l(q) &=& N^*(0) \int {d\theta \over 2\pi } ~{\cos{\theta }~
\cos^2({n~\theta })
\over x -\cos{\theta } + i\eta ~{\rm sgn}(x)}
\nonumber \\
\chi^{(n)}_t(q) &=& N^*(0) \int {d\theta \over 2\pi } ~{\cos{\theta }
{}~\sin^2({n~\theta })
\over x -\cos{\theta } + i\eta ~{\rm sgn}(x)} ~ .
\end{eqnarray}
The function $\beta({\bf S;q})$ in Eq. (\ref{fn}) denotes the angle
between ${\bf S}$ and ${\bf q}$.
In three dimension, the inclusion of $l$th Landau interaction can be
performed using the addition theorem.  The boson propagator is then:
\begin{eqnarray}
\langle a({\bf S};q)~a^{\dag}({\bf S};q)\rangle &=&
{i\over \omega - v_F^* {\bf q\cdot \hat{n}_S}+i\eta ~{\rm sgn}(\omega )}
\nonumber \\
&+&i{2\Lambda ^2\over (2\pi)^3}~{{\bf q\cdot \hat{n}_S}\over \big{[}
\omega - v_F^* {\bf q\cdot \hat{n}_S}+i\eta ~{\rm sgn}(\omega )\big{]}^2}
\sum_{m}{ {\rm Y}^{-m}_l\big{(} {\bf \Omega}({\bf S;q})\big{)}~
{\rm Y}^m_l\big{(} {\bf \Omega}({\bf S;q})\big{)}\over
{(-1)^m\over 4\pi ~f_l}-\chi_l^m (q) }
\end{eqnarray}
where
\begin{eqnarray}
\chi_l^m(q) ={N^*(0)\over 4\pi}\int d^2 \Omega~ {\cos\theta \over
x-\cos \theta + i\eta ~{\rm sgn}(x)}~{\rm Y}^{-m}_l({\bf \Omega})~
{\rm Y}^m_l({\bf \Omega})
\end{eqnarray}
Here, ${\bf \Omega}({\bf S;q})$ denotes the angles between ${\bf S}$ and
${\bf q}$ in spherical coordinates.

In general, when more than one Landau parameter
is involved, we introduce the appropriate number of mediating fields for
each Landau parameter.  For example, in
two dimensions we introduce two mediating fields for a
single Landau interaction $f_n$ as above.  In three dimension, on
the other hand, we decompose the Landau interaction as
\begin{equation}
(2l+1)~f_l~{\rm P}_l(\cos \theta)
=4\pi~f_l \sum_{m=-l}^l (-1)^m~{\rm Y}^m_l\big{(} {\bf \Omega}({\bf S;q})
\big{)}~{\rm Y}^{-m}_l\big{(} {\bf \Omega}({\bf S;q})\big{)}
\end{equation}
and introduce a mediating field $A^m_l(q)$ for each spherical harmonic,
amounting to $2l+1$ fields for each Landau interaction.
The propagators of the mediating fields are then matrices.
Upon the inclusion of gauge interactions along with the Fermi Liquid
interactions, we first construct a gauge-covariant form for the
Fermi liquid interaction and then treat the
gauge fields and the mediating fields on the same footing.
An illustrative calculation is presented in Appendix C for the case of the
Chern-Simons gauge interaction plus the Fermi Liquid parameter $f_1$.

\section{How Kohn's Theorem Is Satisfied}

In this appendix, we prove the result quoted in Sec. \ref{subsec:hfllbose}
that the collective excitations of the interacting quasiparticles of
the half-filled Landau level occur at a frequency $\omega = 2\pi \rho_f
\tilde{\phi}/m$, the cyclotron frequency of a particle with bare mass
$m$ in a magnetic field $B=2\pi \rho_f c\tilde{\phi}/e$.
To prove this, we add to the action of the Chern-Simons gauge theory a
Fermi liquid interaction in which all the coefficients except $f_1$
have been set equal to zero. The contribution to the action
Eq. (\ref{f1}) is given in real space and time by
\begin{equation}
S_{FL} = {f_1\over 2k_F^2}\int d^2x~dt~\psi^{\dag}{\bf \nabla }
\psi \cdot \psi^{\dag}{\bf \nabla }\psi
\end{equation}
which is made gauge covariant by replacing
${\bf \nabla }$ by ${\bf \nabla }-i{\bf A}$. On Fourier transforming
it is given in terms of the currents as
\begin{eqnarray}
S_{FL}
&\approx &
 -{{f_1}\over{2 V k_F^2}}~ \int {{d\omega}\over{2\pi}}~\big{\{}
\sum_{S,T,{\bf q}}~ J(S; q)~ {\bf k_{\it S} \cdot k_{\it T}}~ J(T; -q)
\nonumber \\
 & & +{{f_1}\over{2 V k_F^2}}~
\sum_{S,\bf q} \rho_f \big{[} J(S;q)~{\bf k_{\it S}\cdot A}(-q)
+ J(S;-q)~{\bf k_{\it S}\cdot A}(q) \big{]}
 -{{f_1}\over{2 V k_F^2}}~ \sum_{\bf q}
\rho ^2_f~{\bf A}(q)\cdot {\bf A}(-q) \big{\}}
\end{eqnarray}
As in Sec. \ref{sec:longit} the term quadratic in $J$ is decoupled by
introducing additional transverse $A_t$ and longitudinal $A_l$
fictitious fields.

Including the Fermi liquid interaction the complete action of the
theory is given by
\begin{eqnarray}
S[A^{\mu },a,\xi ,\xi^*] &=& \sum_{ S} \sum_{\bf q, q\cdot \hat{n}_{\it S} >0}
\int {d\omega \over 2\pi}~\big{\{}
(\omega - v_F^*~ {\bf q\cdot \hat{n}_{\it S}})~a^{*}({ S}; q)~a({ S}; q)
+\xi(S;q)~a^*(S;q)+\xi^*(S;q)~a(S;q)\big{\}}
\nonumber \\
&+&{1\over V}\sum_{ S} \sum_{\bf q} \int {d\omega \over 2\pi }~
J(S;q)\Big{\{} A_0(-q)+(v_F^* +{f_1\rho_f \over k_f}){\bf \hat{n}_{\it S}\times
q\over |q|}~A_T(-q)
\nonumber \\
&+&{\bf \hat{n}_{\it S} \cdot q\over |q|}~A_l(-q)+
{\bf \hat{n}_{\it S} \times q\over |q|}~A_t(-q)\Big{\}}
\nonumber \\
&+&{1\over 2V}\sum_{\bf q} \int {d\omega \over 2\pi }~\Big{\{}
\big{[}{\rho_f \over m^*}+
{f_1\rho_f^2 \over k_F^2} + {{{\bf q}^2~ V({\bf q})}
\over{(2 \pi \tilde{\phi})^2}} \big{]}~A_T(q)~
A_T(-q)+ {2i|{\bf q}|\over 2\pi \tilde{\phi}}~A_0(q)~A_T(-q)
\nonumber \\
&-&{1\over f_1}~ A_l(q)~A_l(-q)-{1\over f_1}~A_t(q)~A_t(-q) \Big{\}}~ ,
\label{CSFL}
\end{eqnarray}
where $\mu $ runs over $0, T, l$ and $t$.
and we couple the boson fields to $\xi , \xi^* $ to construct a generating
functional as in Eq.(\ref{ftnal}). After integrating out the
boson fields $a, a^*$, we obtain an effective action:
\begin{eqnarray}
S_{eff}[A^{\mu },\xi ,\xi^* ] &=&
{1\over 2}\int {d^2q\over (2\pi )^2} \int {d\omega \over 2\pi }~\Big{\{ }
\big{[} {\rho_f \over m^*}+{f_1\rho_f^2 \over k_F^2}+ {{{\bf q}^2~V({\bf q})}
\over{(2 \pi \tilde{\phi})^2}}
+(v_F^*+{f_1\rho_f \over k_F})^2\chi_t(q)\big{]}~A_T(q)~A_T(-q) \nonumber \\
&+& {2i|{\bf q}|\over 2\pi \tilde{\phi}}~A_0(q)~A_T(-q)+[-{1\over f_1}
+\chi_l(q)]~A_l(q)~A_l(-q)+[-{1\over f_1}+\chi_t(q) ]
{}~A_t(q)~A_t(-q)
\nonumber \\
&+&\chi_{l0}(q)~[A_l(q)~A_0(-q)-A_0(q)~A_l(-q)]
+2[v_F^*+{f_1\rho_f \over k_F}]~\chi_t(q)~A_T(q)~A_t(-q)
\nonumber \\
&+& \chi ^0(q)~ A_0(q)~ A_0(-q) \Big{\} }
- {1\over V}\sum_{\bf q,q\cdot \hat{n}_{\it S} >0} \int {d\omega \over 2\pi }~
\sum_S \Big{\{}{\xi(S;q) \sqrt{\Omega |{\bf q\cdot \hat{n}_{\it S}}|}\over
(\omega -v_F^*{\bf q\cdot \hat{n}_{\it S} })}
{}~A_{\alpha }(-q)~\epsilon^{\alpha }(
S;q) \nonumber \\
&+& {\xi^*(S;q) \sqrt{\Omega |{\bf q\cdot \hat{n}_{\it S}}|}\over
(\omega -v_F^*{\bf q\cdot \hat{n}_{\it S}})}~A_{\alpha }(q)~\epsilon^{\alpha }(
S;-q)
+{\xi^*(S;q)~\xi(S;q)\over \omega -v_F^*{\bf q\cdot \hat{n}_{\it S}}}
\Big{\}} ~ ,
\label{AppSeff}
\end{eqnarray}
here
\begin{equation}
\epsilon^{\alpha }(S;q) = (1, [v_F^*+{f_1\rho_f \over k_F}]{\bf \hat{n}_{\it S}
\times q \over |q|},{\bf \hat{n}_{\it S} \cdot  q \over |q|},
{\bf \hat{n}_{\it S} \times q \over |q|})
\end{equation}
and the susceptibilities $\chi_{\alpha }(q)$ are defined by:
\begin{eqnarray}
\chi^0(q) &=& -N^*(0)\int {d\theta \over 2\pi } ~{\cos{ \theta }\over x -
\cos{\theta } + i\eta ~{\rm sgn}(x) }\nonumber \\
&=& N^*(0)\big{[}1-\theta(x^2-1){|x|\over \sqrt{x^2-1} }
+i\theta(1-x^2){|x|\over \sqrt{1-x^2} }\big{]} \nonumber \\
\chi_t(q) &=& N^*(0)\int {d\theta \over 2\pi } ~{\cos{\theta }~\sin^2{\theta }
\over x -\cos{\theta } + i\eta ~{\rm sgn}(x)} \nonumber \\
&=& N^*(0)\big{[}x^2-{1\over 2}-\theta(x^2-1)|x| \sqrt{x^2-1}
-i\theta(1-x^2)|x| \sqrt{1-x^2} \big{]} \nonumber \\
\chi_l(q) &=& N^*(0)\int {d\theta \over 2\pi } ~{\cos^3{\theta }
\over x -\cos{\theta } + i\eta ~{\rm sgn}(x)} \nonumber \\
&=& N^*(0)\big{[}-x^2-{1\over 2}+\theta(x^2-1)|x|{x^2\over \sqrt{x^2-1} }
-i\theta(1-x^2)|x|{x^2\over \sqrt{1-x^2} }\big{]} \nonumber \\
\chi_{l0}(q)&=&-\chi_{l0}(-q)=N^*(0)\int {d\theta \over 2\pi }~{\cos^2{\theta }
\over x -\cos{\theta } + i\eta ~{\rm sgn}(x)} \nonumber \\
&=& N^*(0)\big{[}-x+\theta(x^2-1)|x|{x\over \sqrt{x^2-1} }
-i\theta(1-x^2)|x|{x\over \sqrt{1-x^2} }\big{]} ~ ,
\label{chis}
\end{eqnarray}
and $x=\omega /(v_F^* |{\bf q}|)$.
The boson propagator is obtained in the usual way
\begin{eqnarray}
\langle a(S; q)~ a^\dagger( S; q)\rangle
&=& {i\over \omega - v_F^* {\bf q\cdot \hat{n}}_S +i \eta~ {\rm sgn}(\omega)}
+ i~ {\Lambda \over (2\pi )^2}~ {\bf q\cdot \hat{n}}_S~
{D_{\alpha \beta }(q)~\epsilon ^\alpha (S;q)
{}~\epsilon ^\beta (S;-q)\over
[\omega - v_F^* {\bf q\cdot \hat{n}}_S + i \eta ~{\rm sgn}(\omega)]^2} ~ ,
\label{AppGaa}
\end{eqnarray}
and the collective modes are given by the poles of the gauge propagator.
The inverse gauge propagator can be read off from the action
Eq. (\ref{AppSeff}) and in the limit of interest
$\omega \gg  v_F^*|{\bf q}|$ is given by
\begin{equation}
[K(q)]_{\alpha \beta} =
\left( \begin{array}{cccc}
-{N^*(0)\over 2x^2} & {i|{\bf q }|\over 2\pi \tilde{\phi }} & -{N^*(0)
\over 2x} & 0
\\
{i|{\bf q }|\over 2\pi \tilde{\phi }}
& \rho_f({1\over m^*}+{f_1\rho_f\over k_F^2
}) & 0 & -{N^*(0)\over 8x^2}(1+{m^*f_1\rho_f \over k_F^2})
\\
{N^*(0)\over 2x} & 0 & -{1\over f_1} & 0
\\
 0 & -{N^*(0)\over 8x^2}(1+{m^*f_1\rho_f \over k_F^2}) & 0 & -{1\over f_1}
\end{array} \right)
\end{equation}
The poles of $D_{\alpha \beta}$
are found by setting ${\rm det}|K_{\alpha \beta }| = 0$ which gives
$\omega ^2 = [2\pi \tilde{\phi} \rho_f ({1 / m^*}+{f_1 / 4\pi })]^2
= (2\pi \rho_f \tilde{\phi} / m_b)^2$.  The
second equality follows from the application of Galilean invariance to a
Fermi liquid\cite{Tony}.  The
collective mode appears at the cyclotron frequency determined by the bare
electron band mass in agreement with Kohn's theorem.

\begin{figure}
\caption{The first of two 1/N vertex correction, $\Lambda^{(1)}(p; q)$.
The symbol {\bf X} indicates that external lines are to be amputated.}
\label{vertex1}
\end{figure}

\begin{figure}
\caption{The two diagrams which give the second 1/N vertex corrections,
$\Lambda^{(2)}(p; q)$.  The leading
divergences in these two diagrams cancel via Furry's theorem (see text)
as only the nonlinear
terms in the fermion dispersion break charge-conjugation symmetry.}
\label{vertex2}
\end{figure}

\begin{figure}
\caption{Spectral function ${\rm Im} G(k, \omega)$ in two dimensions
for the case $F_0 = 2$.  Slices at three different
frequencies $\omega = 0, k_F/8$, and $k_F/4$ are shown respectively in
(a), (b) and (c).
Here $\lambda = k_F$, $v_F^* = 1$, and a lattice of
$2048 \times 2048$ points is used to compute the Fourier transforms.  The
peak has been centered in each plot around the point $k = \omega$.}
\label{spectral1}
\end{figure}


\begin{references}
\bibitem{Halperin}B. I. Halperin, P. A. Lee and N. Read,  Phys. Rev. B
{\bf 47}, 7312 (1993).
\bibitem{Furusaki}Yong Baek Kim, Akira Furusaki, Xiao-Gang Wen, and
Patrick Lee, ``Gauge-invariant response functions of fermions coupled to a
gauge field,''  to appear in Phys. Rev. B.
\bibitem{Altshuler}B. L. Altshuler, L. B. Ioffe, and A. J. Millis,
Phys. Rev. B {\bf 50}, 14048 (1994).
\bibitem{Shankar}R. Shankar,  Physica A {\bf 177}, 530 (1991);
 Rev. Mod. Phys. {\bf 66}, 129 (1994).
\bibitem{Gan}J. Gan and E. Wong,  Phys. Rev. Lett. {\bf 71}, 4226 (1993).
\bibitem{Nayak}Chetan Nayak and Frank Wilczek, Nucl. Phys. B {\bf 417}, 359
(1994);  Princeton University preprint (cond-mat/9408016).
\bibitem{Khvesh}D. V. Khveshchenko and P. C. E. Stamp,  Phys. Rev. Lett.
{\bf 71}, 2118 (1993);  Phys. Rev. B {\bf 49}, 5227 (1994).
\bibitem{Castellani}C. Castellani, C. Di Castro  and W. Metzner,  Phys.
Rev. Lett. {\bf 72}, 316 (1994); C. Di Castro, C. Castellani, and W. Metzner,
``Conservation laws in normal metals: Luttinger liquid vs. Fermi liquid,''
in {\it The Physics and the Mathematical
Physics of the Hubbard Model} edited by
D. Campbell (Plenum Press, New York, in press); C. Castellani and C. Di Castro,
``Crossover from Luttinger to Fermi liquid by increasing dimension,''
in {\it Proceedings of IV International Conference on Materials
and Mechanisms of Superconductivity, Grenoble, July 1994,}
edited by P. Wyder (Elsevier Publications, 1995).
\bibitem{Haldane}F. D. M. Haldane, ``Luttinger's Theorem and Bosonization
of the Fermi Surface,'' in {\it
 Proceedings of the International School of Physics
``Enrico Fermi,'' {\bf 121} Varenna 1992} edited by R. Schrieffer and
R. A. Broglia (North-Holland, New York, NY 1994).
\bibitem{Tony}A. Houghton and J. B. Marston,  Phys. Rev. B {\bf 48},
7790 (1993).
\bibitem{HKM}A. Houghton, H.-J. Kwon, and J. B. Marston,
 Physical Review B {\bf 50}, 1351 (1994).
\bibitem{HKMS}A. Houghton, H.-J. Kwon, J. B. Marston and R. Shankar,
 J. Phys.: Cond. Matt. {\bf 6}, 4909 (1994).
\bibitem{KHM}H.-J. Kwon, A. Houghton, and J. B. Marston,  Phys. Rev. Lett
{\bf 73} 284 (1994).
\bibitem{Castro}A. H. Castro-Neto and Eduardo H. Fradkin,
 Phys. Rev. Lett {\bf 72}, 1393 (1994);  Phys. Rev. B {\bf 49},
10877 (1994).
\bibitem{Castro2}A. H. Castro-Neto and Eduardo H. Fradkin,
``Exact solution of the Landau fixed points via bosonization,''
UIUC preprint (1994).
\bibitem{Kopietz}Peter Kopietz and Kurt Schonhammer, ``Functional
bosonization of interacting fermions in arbitrary dimensions,'' Gottingen
preprint (1994).
\bibitem{DVK}D. V. Khveshchenko, Phys. Rev. B {\bf 49}, 16893
(1994); ``Geometrical approach to bosonization of
$D > 1$ dimensional (non) Fermi liquids,'' Princeton preprint (1994).
\bibitem{Gauge}See, for example,
G. Baskaran and P. W. Anderson,  Phys. Rev. B {\bf 37},
580 (1988); I. Affleck and J. B. Marston,  Phys. Rev. B {\bf 37}, 3774
(1988); L. B. Ioffe and A. I. Larkin,  Phys. Rev. B {\bf 39},
8988 (1989).
\bibitem{Lee}P. A. Lee and N. Nagaosa,  Phys. Rev. B {\bf 46},
5621 (1992).
\bibitem{Blok}B. Blok and H. Monien,  Phys. Rev. B {\bf 47}, 3454 (1993).
\bibitem{Kalmeyer}V. Kalmeyer and S. C. Zhang,  Phys. Rev. B
{\bf 46}, 9889 (1992).
\bibitem{Kohn}W. Kohn,  Phys. Rev. {\bf 123}, 1242 (1961).
\bibitem{Hertz}J. A. Hertz, Phys. Rev. B {\bf 14}, 1165 (1976).
\bibitem{Luther}A. Luther,  Phys. Rev. B {\bf 19}, 320 (1979).
\bibitem{Wen}P. Bares and X. G. Wen,  Phys. Rev. B {\bf 48}, 8636 (1993).
\bibitem{Jiang}H.W. Jiang {\it et al.}  Phys. Rev. B {\bf 40},
12013 (1989); R. L. Willet {\it et al.}  Phys. Rev. Lett. {\bf 65}, 112
(1990); R. R. Du {\it et al.},  Phys. Rev. Lett. {\bf 70}, 2944 (1993);
R. L. Willett {\it et al.}  Phys. Rev. B {\bf 47}, 7344 (1993);
W. Kang {\it et al.}  Phys. Rev. Lett. {\bf 71} 3850 (1993);
H. C. Manoharan, M. Shayegan, and S. J. Klepper,  Phys. Rev. Lett.
{\bf 73}, 3270 (1994); R. R. Du {\it et al.},  Phys. Rev. Lett. {\bf 73},
3274 (1994).
\bibitem{Willett}R. L. Willett, R. R. Ruel, K. W. West, and L. N. Pfeiffer,
 Phys. Rev. Lett. {\bf 71}, 3846 (1993).
\bibitem{Du}R. R. Du {\it et al.}  Solid State Comm. {\bf 90}, 71 (1994).
\bibitem{Leadley}D. R. Leadley, R. J. Nicholas, C. T. Foxon, and
J. J. Harris,  Phys. Rev. Lett. {\bf 72}, 1906 (1994).
\bibitem{Goldman}V. J. Goldman, B. Su, and J. K. Jain,  Phys. Rev. Lett.
{\bf 72}, 2065 (1994).
\bibitem{Lopez}A. Lopez and E. Fradkin,  Phys. Rev. B {\bf 44},
5246 (1991);  Phys. Rev. Lett. {\bf 69}, 2126 (1992).
\bibitem{Kim}Yong Baek Kim and Xiao-Gang Wen,
Phys. Rev. B {\bf 50}, 8078 (1994).
\bibitem{Simon}Steven H. Simon and Bertrand I. Halperin,  Phys. Rev. B
{\bf 48}, 17368 (1993)
\bibitem{old}F. D. M. Haldane,  J. Phys. C {\bf 14}, 2585 (1981).
\bibitem{Furry}See, for example, Claude
Itzykson and Jean-Bernard Zuber, {\it Quantum Field Theory}
(McGraw-Hill, Inc., New York, 1980), p. 276.
\bibitem{IKL}L. B. Ioffe, S. Kivelson, and A. I. Larkin,  Phys. Rev. B
{\bf 44}, 12,537 (1991).
\bibitem{Kohna}W. Kohn,  Phys. Rev. Lett. {\bf 2}, 393 (1959).
\bibitem{Peschel}A. Luther and I. Peschel,  Phys. Rev. B {\bf 9},
2911 (1974).
\bibitem{Stern}F. Stern,  Phys. Rev. Lett. {\bf 18}, 546 (1967).
\bibitem{Pethick}C. J. Pethick, G. Baym and H. Monien,  Nuclear Physics
A{\bf 498}, 313c (1989).
\bibitem{Norton}Sudip Chakravarty, Richard E. Norton, and Olav F.
Syljuasen, ``Transverse gauge interactions and the vanquished Fermi liquid,''
UCLA preprint (1994).
\bibitem{Jain}J. K. Jain,  Phys. Rev. Lett. {\bf 63}, 199 (1989).
\bibitem{KLWS}Yong Baek Kim, Patrick A. Lee, Xiao-Gang Wen, and P. C. E.
Stamp, ``Influence of gauge-field fluctuations on composite fermions near
the half-filled state,'' MIT preprint (1994).
\bibitem{code}A copy of the ``C''-code executing these operations is available
from the authors.

\end{references}
\end{document}